\documentclass[%
 reprint,
superscriptaddress,
nofootinbib,
 amsmath,amssymb,
 aps,
 pra,
]{revtex4-1}
\usepackage{verbatim}
\usepackage{mathtools}
\usepackage{amsthm}
\usepackage{physics}
\usepackage{graphicx}
\usepackage{dcolumn}
\usepackage{bm}
\usepackage{xcolor}
\usepackage{hyperref}
\usepackage{xfrac}
\usepackage{placeins}
\graphicspath{ {./images_labelled/} }

\begin{document}

\title{Structure of correlations for the single-$j$ model using quantum information theory}

\author{Javier Faba}

\affiliation{%
Departamento de Arquitectura y Tecnología de Sistemas Informáticos, Universidad Polit\'ecnica de Madrid, Campus Montegancedo, 28660 Boadilla del Monte, Madrid, Spain
}%
\affiliation{%
Center for Computational Simulation, Universidad Polit\'ecnica de Madrid, Campus Montegancedo, 28660 Boadilla del Monte, Madrid, Spain
}%
\author{Vicente Mart\'\i n}

\affiliation{%
Departamento de Lenguajes y Sistemas Informáticos e Ingeniería de Software, Universidad Polit\'ecnica de Madrid, Campus Montegancedo, 28660 Boadilla del Monte, Madrid, Spain
}%
\affiliation{%
Center for Computational Simulation, Universidad Polit\'ecnica de Madrid, Campus Montegancedo, 28660 Boadilla del Monte, Madrid, Spain
}%

\author{Luis Robledo}
\affiliation{%
Departamento de F\'isica Te\'orica and CIAFF, Universidad Aut\'onoma de Madrid, E-28049 Madrid, Spain 
}%
\affiliation{%
Center for Computational Simulation, Universidad Polit\'ecnica de Madrid, Campus Montegancedo, 28660 Boadilla del Monte, Madrid, Spain
}%


\begin{abstract}

We say that a system undergoes a quantum phase transition when the mean-field ground state of a given Hamiltonian spontaneously breaks its symmetries depending on the Hamiltonian parameters. It is a widely accepted view that this phenomenon enables the mean-field state to absorb the maximum amount of correlations without losing its mean-field structure. However, it is still uncertain how those correlations are incorporated within the mean-field state and how the correlation structure changes when we get closer to the exact ground state by restoring the symmetries. In this work, we answer these questions by studying the single-$j$ model, which is a realistic fermionic nuclear model with abelian and non-abelian broken symmetries. We find that both types of symmetries have very different effects for the correlations within the mean-field state. This makes unclear the concept of the symmetry breaking process as a mechanism to absorb the maximum amount of correlations. Moreover, the symmetry restoration redistributes existing `fake' correlations between paired orbitals to more complex subsystems.

\end{abstract}

\pacs{Valid PACS appear here}
\maketitle


\section{\label{sec:introduction}Introduction}

Spontaneous symmetry breaking at the mean field level is a central concept in our present understanding of low energy nuclear structure. Due to the non-linear nature of  mean field approaches (both Hartree-Fock or Hartree-Fock-Bogoliubov) the solution of the mean field equation not necessarily preserve the symmetries of the Hamiltonian. By breaking symmetries of the Hamiltonian at the mean field level one can incorporate relevant correlations into the simple product wave function of the mean field. The price to pay is that the mean-field wave functions cannot be labeled by symmetry quantum numbers like angular momentum or parity and therefore are not a faithful representation of the nuclear wave function \cite{RingSchuck1980}. In order to connect the mean field wave function (referred to as "intrinsic wave function") with the one of a real nucleus (referred to as "laboratory wave function") symmetry restoration procedures must be used. Those procedures are implemented through the use of projection-onto-good-quantum-number methods deeply rooted on the underlying group structure of the symmetry to be restored. Projection techniques are essential to consider many body correlations  in a framework preserving good quantum numbers and, as such, represents the basis of many state-of-the-art beyond-mean-field approaches used nowadays in nuclear physics \cite{Sheikh}.

Although this procedure is remarkably successful, the physical meaning of the correlations generated by symmetry breaking and the subsequent symmetry restoration remains only partially understood. It is commonly argued that symmetry breaking enables the mean-field state to absorb the largest possible amount of correlations compatible with its product-state structure, while symmetry restoration introduces additional correlations that bring the wave function closer to the exact many-body state by restoring proper quantum numbers. However, this qualitative interpretation does not explain how those correlations are actually encoded in the many-body wave function, how they are redistributed by the restoration procedure, or whether different broken symmetries affect the correlation structure in the same way.

Quantum information theory provides a natural framework for addressing these questions. Measures such as entanglement entropy, mutual information, and quantum discord (QD) quantify correlations independently of dynamical considerations and make it possible to characterize the internal structure of many-body quantum states. During the last years, the extension to fermion systems of these quantities \cite{fermionic_entanglement_study,gigena_1,Friis_2013,Friis_2016} has opened the possibility of studying realistic nuclear wave functions from an information-theoretical perspective \cite{robin_2021,faba_1,faba_2}. Previous works have successfully applied these tools to schematic nuclear models with Abelian symmetries, revealing how quantum correlations evolve across quantum phase transitions and under symmetry restoration \cite{faba_3}.

In this work, we extend the analysis of Ref \cite{faba_3} to the single-$j$ (pairing-plus-quadrupole) model, which is a simple but realistic system in which the mean field solution can break both particle number and rotational symmetries depending of the parameters of the Hamiltonian \cite{kisslinger1960ppq,Baranger1965}. This model is particularly suitable for investigating the interplay between pairing and quadrupole correlations, the associated quantum phase transitions, and the different roles played by an Abelian and a non-Abelian symmetry within the same framework. By combining Hartree-Fock-Bogoliubov theory, symmetry restoration techniques, and fermionic quantum information measures, we analyze how correlations are distributed within the system and how this distribution changes after restoring particle number and angular momentum, providing a microscopic interpretation of the correlation structure associated with symmetry breaking and restoration.

This paper is structured as follows. Sec. \ref{sec:theory} is a review of the single-$j$ model and the fermionic quantum information tools that will be used to obtain the results. These are shown in Sec. \ref{sec:results}, and they cover the HFB ground state together with the $J=0$ and full symmetry restoration. Finally, Sec. \ref{sec:conclusions} contains a discussion and conclusions of the results shown and further work is proposed.

\section{\label{sec:theory}Theoretical background}

This section outlines the theoretical framework of our study. In Sec.~\ref{sec:singlej}, we review the single-$j$ model along with the HFB formalism and the symmetry restoration techniques employed. Next, in Sec.~\ref{sec:correlations}, we introduce the fermionic quantum information measures used to quantify classical and quantum correlations among single-particle orbitals.

\subsection{\label{sec:singlej}The single-$j$ model}

The single-$j$ model, also called pairing-plus-quadrupole model, is a realistic yet simple fermionic model for nuclear structure \cite{kisslinger1960ppq}. It assumes that all relevant physical processes within a nucleus occur within a single valence shell with angular momentum $j$, and that the nucleons in this shell interact through pairing and quadrupole forces \cite{Baranger1965}. Those forces represent a simple approximation to the effective residual interaction left after taking into account mean field effects. Bare nucleon-nucleon forces cannot be used at this point because of its strong repulsive character that prevents its use in a plain mean field calculation. After taking into account in-medium effects, mostly associated to Pauli exclusion principle, one can extract an effective interaction adapted for mean field studies. The resulting residual interaction after removal of the monopole terms of the mean field can only be obtained numerically after sophisticated many body techniques. The low energy nuclear phenomenology suggests that the first non trivial term in a multipole-multipole expansion of the force (the one associated to the quadrupole tensor) plus a simple pairing interaction is enough to grasp most of the low energy physics in the atomic nucleus \cite{Dufour96}. This is the heuristic justification of both the pairing plus quadrupole model and its simpler version where only a single $j$ shell is taken as active space. For simplicity, we will assume that all the nucleons are spinless fermions of the same kind. Each single particle state will be labeled by the total angular momentum of the shell, $j$, and the third component, $m$. Hence, the number of single particle states is $2j+1$. We will assume that the energy level of each single particle state is unperturbed, so all single particle states will be degenerated and the corresponding energy will be denoted by $\epsilon$.

Using the pair creation operator,
\begin{equation}
    P^\dagger = \sum_{m} c^\dagger_m c^\dagger_{\overline{m}}
\end{equation}
which creates a superposition of pair of particles within a single particle state and its time reversal conjugate, and the quadrupole operator,
\begin{equation}
    Q_{\mu} = \sum_{mn}q^{\mu}_{mn}c^\dagger_m c_n
\end{equation}
where $q^{\mu}_{mn} = \bra{j,m} Q_\mu\ket{j,n}$ and $\mu \in [-2,-1,0,1,2]$, we can write the single-$j$ Hamiltonian as
\begin{equation}
    H = \epsilon\sum_{m}c^\dagger_m c_m - GP^\dagger P - \chi Q\cdot Q
\end{equation}
where $\epsilon$ is the single particle energy, $G$ and $\chi$ are parameters that quantify the strength of the pairing and quadrupole interaction respectively, and the scalar product of the last term is defined in terms of spherical coordinates, that is, $Q\cdot Q = \sum_\mu (-1)^\mu Q_\mu Q_{-\mu}$. Since the products $P^\dagger P$ and $Q\cdot Q$ are scalar operators, this Hamiltonian is rotational invariant. Also, since all terms are particle conserving, the particle number symmetry is preserved. Depending on the values of $\chi$ and $G$, the approximate ground state solution corresponding to the quasi-particle vacuum may break some of those symmetries, giving rise to a rich phase diagram \cite{dobaczewski_1,vallieres_1,Baranger1965}.

\subsubsection{\label{sec:symmetry_restoration}Mean field solution and symmetry restoration}

The model's interaction contains a pairing term that favors the creation of Cooper pairs and therefore the mean-field treatment of the problem must involve the well-known Hartree-Fock-Bogoliubov (HFB) method \cite{RingSchuck1980}, as it is customary in the implementation of the mean field method in nuclear physics. The quasiparticle HFB creation and annihilation operators are expanded as a linear combination of creation and annihilation operators of the  single particle basis in terms of the Bogoliubov amplitudes $U$ and $V$ of the canonical transformation
\begin{align}  \beta_{k}&=\sum_{l}U_{lk}^{*}c_{l}+V_{lk}^{*}c_{l}^{+}\\\beta_{k}^{+}&=\sum_{l}U_{lk}c_{l}^{+}+V_{lk}c_{l}     
\end{align}
The $U$ and $V$ amplitudes are not linearly independent as the transformation must preserve canonical anticommutation relations (CAR), implying that the block matrix
\begin{equation}
W=\left(\begin{array}{cc}
U & V\\
V & U^{*}
\end{array}^{*}\right)    
\end{equation}
must be unitary $WW^{+}=W^{+}W=\mathbb{I} $. The HFB ground state $|\phi\rangle$ is defined as the lowest energy state annihilated by all  quasiparticle operators $\beta_\mu$ and is simply given by $\ket{\phi}=\prod_{k}\beta_{k}\ket{-}$, where $\ket{-}$ is the bare vacuum. One of the most salient characteristics of the method is the possibility to break particle number symmetry, meaning that $|\phi\rangle$ is in general a linear combination of wave functions each of them having different number of particles. However, the Parity Superselection Rule (PSSR) prevents the mixing of wave functions with even or odd number of particles \cite{Friis_2016}. Depending on the strength parameters of the quadrupole and pairing interaction term, the ground state solution of the single-$j$ model  can break the two symmetries of the Hamiltonian: particle number symmetry and rotational symmetry. This spontaneous symmetry-breaking mechanism is typically considered as a clever artifact of the mean field that allows the incorporation of different types of correlations whithin the simple mean field scheme. However, the exact eigenstates of the Hamiltonian should have quantum numbers corresponding with the Hamiltonian's symmetries, and therefore a symmetry restoration procedure applied to the correlated mean-field solutions is mandatory. In the present model, one has to restore both the particle number and the rotational invariance.

To recover states with well defined quantum numbers, in this case, particle number $N$ and angular momentum $J=0$, Projection-After-Variation (PAV) symmetry restoration is applied to the symmetry-broken HFB state $|\phi\rangle$ \cite{RingSchuck1980,Sheikh}. The particle number projection operator $\hat{P}^N$ and the $J=0$ angular momentum projection operator $\hat{P}^{(J=0)}$ are defined as:
\begin{equation}
\hat{P}^{(N)}=\frac{1}{2\pi} \int_0^{2\pi} e^{i\phi(\hat{N}-N)}
\end{equation}
\begin{equation}
\label{eq:P_operator_J0}
\hat{P}^{(J=0)} = \frac{1}{8\pi^2} \int d\Omega \hat{R}(\Omega),
\end{equation}
where $\hat{N}$ is the particle number operator and $\hat{R}(\Omega) = e^{-i\alpha \hat{J}_z} e^{-i\beta \hat{J}_y} e^{-i\gamma \hat{J}_z}$ represents the spatial rotation operator for Euler angles $\Omega = (\alpha, \beta, \gamma)$, and $\hat{J}_k$ with $k\in\{x,y,z\}$ is the $k$-component of the angular momentum operator. The computation of the one-body density matrix, pairing tensor and two-body density matrix for the symmetry restored states requires the evaluation of matrix elements and overlaps between original and rotated mean-field states. Those quantities are computed through the generalized Wick's theorem \cite{RingSchuck1980,Bertsch12} and Robledo's Pfaffian formula for the norm overlaps \cite{robledo_2009}, respectively. One of the most striking  consequences of the $J=0$ angular momentum restoration is that the symmetry restored one-body density matrix becomes proportional to the identity, and the natural orbitals are given by the single particle $(j,m)$ states \cite{faba_thesis}.

\subsection{\label{sec:correlations}Correlations in fermion systems}

Quantum information tools allow us to quantify, in a very precise way, different types of correlations within a quantum state. The concepts of entanglement entropy, negativity, mutual information \cite{Nielsen_Chuang_2010} or quantum discord \cite{Henderson_2001,Ollivier_2001} are well defined and allow us to analyze the quantum and classical correlation structure of a quantum system. However, those measures require to define the partial trace operation and, at the same time, this requires to define a tensor product structure within our total Hilbert space. While for distinguishable systems (e.g., qubits) this is straightforward, it is not for indistinguishable ones, which includes fermionic systems. The impossibility of decomposing the Fock space directly into a tensor product makes harder to define fermionic quantum information measures, since there is no consistent mapping between qubits and fermions \cite{Friis_2016}. More precisely, it is necessary to define a specific fermionic partial trace operation together with the PSSR to obtain meaningful quantum information measures \cite{Friis_2016}. In this sense, we need to define the single particle states as the mathematical objects that define the locality of the system \cite{gigena_1,gigena_2,fermionic_entanglement_study,Friis_2013,bannuls_2007}. Hence, all fermionic quantum information measures that we will use throughout this work measure correlations among single particle states, also called orbitals.

We will use fermionic quantum information tools to characterize in a precise and consistent way the correlation structure of the mean-field and symmetry restored fermionic ground states of the single-$j$ model. Following previous work \cite{faba_1,faba_2,faba_3}, we will use the fermionic quantum discord between orbital $A$ and orbital $B$, defined as
\begin{equation}
\label{eq:quantum_discord}
    \delta(A,B) \equiv I(A,B)-J(A,B)
\end{equation}
where the mutual information and the classical correlations between the $A$ and $B$ orbitals are defined as
\begin{equation*}
\label{eq:mutual_information_quantum}
    I(A,B) = S(\rho^{(A)})+S(\rho^{(B)})-S(\rho^{(A,B)})
\end{equation*}
\begin{equation*}
\label{eq:classical_correlations}
    J(A,B) \equiv \max_{\{\Pi_k^{(B)}\}}S(\rho^{(A)})-S(\rho^{(A,B)}|\{\Pi_k^{(B)}\})
\end{equation*}
respectively. Here $S(\rho^{(X)})$ stands for the von Neumann entropy of the reduced state corresponding to the subsystem $X$, and the quantum conditional entropy is
\begin{equation}
\label{eq:quantum_conditional_entropy}
    S(\rho^{(A,B)}|\{\Pi_k^{(B)}\}) \equiv \sum_k p_kS(\rho^{(A,B)}_k)
\end{equation}
where $\{\Pi_k^{(B)}\}$ is a complete set of orthogonal projectors within the subspace corresponding to the orbital $B$, $\rho_k^{(A,B)} = \frac{1}{p_k}\Pi_k^{(B)}\rho^{(A,B)}\Pi_k^{(B)}$ is the measured projected state and $p_k = \Tr(\Pi_k^{(B)}\rho^{(A,B)}\Pi_k^{(B)})$ is the associated probability \cite{faba_1}. The two-orbital fermionic quantum discord in Eq. (\ref{eq:quantum_discord}) can be used as a measure of the total quantum correlations (including entanglement and beyond) between two subsystems. Taking into account the PSSR, this quantity is extremely easy to compute and can be applied to pure and mixed states, as shown in \cite{faba_1}. As we will see, this will be useful for the $J=0$ symmetry restoration, since the reduced state corresponding to an orbital pair is not pure anymore. Additionally, since the total correlations are given by the mutual information, those two quantities allow us to distinguish between quantum and classical correlations. Moreover, since all ground states are pure quantum states, we can use the fermionic partial trace to obtain the one-body entropy, $S(\gamma)$\footnote{The one-body density matrix is denoted by $\gamma$, see Appendix \ref{sec:AppendixA} for its definition.}, and the two-orbital entropy, $S(\rho^{(A,B)})$, which are measures of the total correlations within the whole system \cite{gigena_1} and between two orbitals and the rest, respectively.

All of these measures are of special interest if we take into account the structure of the mean-field HFB state, since we will be able to quantify:
\begin{itemize}
    \item Classical and quantum correlations between paired orbitals through mutual information and quantum discord applied to paired orbitals. This allows us to characterize the amount of correlations within an orbital pair. This is interesting since, as we will see, following the HFB ansatz, this is the only possible source of correlation when dealing with natural orbitals.
    \item Total correlations between an orbital pair and the rest through the two-orbital entropy. As it is shown in Appendix \ref{sec:AppendixA}, the reduced state corresponding to two paired natural orbitals for the HFB ansatz is always pure. This implies that there is no correlation between paired natural orbitals and the rest. However, this is not generally true for the symmetry restored state, giving rise to a new source of correlation beyond mean-field.
    \item Amount of total correlations within the whole system through the one-body entropy. Typically, the amount of correlations within a system has been quantified through the correlation energy. However, it has been shown in \cite{faba_2} that this quantity is not a good tool to quantify them, in contrast to the one-body entropy.
\end{itemize}

\section{\label{sec:results}Results}

In this section, we analyze the correlation structure of the single-$j$ model across different levels of approximation using the quantum information tools described above. We begin in Sec.~\ref{sec:results_corr_measures} by examining how spatial rotations and gauge transformations affect correlation measures, establishing the natural orbital basis as the appropriate framework to study intrinsic correlations. In Sec.~\ref{sec:results_corr_structure_HFB}, we investigate the correlations encoded in the symmetry-broken HFB ground state across the quantum phase transition. Finally, in Secs.~\ref{sec:results_corr_structure_J0} and \ref{sec:results_corr_structure_NJ0}, we evaluate how these correlations are redistributed when restoring angular momentum ($J=0$) and both particle number and angular momentum ($N, J=0$), respectively.

\subsection{\label{sec:results_corr_measures}Correlation measures and symmetry transformations}

In general, when the Hamiltonian $H$ has some symmetry, there exists a family of unitary operators $\{U_i\}$ that represent the corresponding symmetry transformations and commute with the Hamiltonian. It is the case, for example, of rotations in the three dimensional space. The unitary operators representing all possible rotations in the three dimensional space, $\{R(\Omega)\}$, are labelled by the Euler angles, denoted by $\Omega$, and $[H,R(\Omega)] = 0$ for all $\Omega$ if the Hamiltonian is invariant under rotations. Given an arbitrary state $\ket{\psi}$, it is easy to check that all the states within the family of transformed states $\ket{\psi_i} = U_i\ket{\psi}$ give the same expectation value for the energy, this is, $\bra{\psi_i}H\ket{\psi_i} = \bra{\psi}H\ket{\psi}$. In this sense, we say that the states $\ket{\psi_i}$ are degenerated\footnote{This is actually a language abuse, since in general the state $\ket{\psi}$ is not a eigenstate of $H$.}.

When we solve numerically the variational problem to find the HFB ground state for a Hamiltonian $H$, we need to define an initial HFB state to initiate the minimization process. The choice is rather arbitrary, and the optimal one (with respect to the computational cost and convergence of the numerical method) depends on the computational details of the minimization process  and particular properties of the model. If the variational method is successful, the energy will be always the lowest possible regardless of the choice of the initial state. However, due to the symmetries of the Hamiltonian, the final state can be any degenerated state of the family $\{\ket{\psi_i}\}$. Which state is the final one depends on the computational details of the variational method and the initial HFB state. For example, if the Hamiltonian is invariant under rotations, and the resulting HFB ground state spontaneously breaks the symmetry (i.e., it is deformed) the orientation of the matter distribution can point into any arbitrary orientation without affecting the energy. Hence, if we start the variational process with a random HFB state and we use the gradient descent as the variational optimization algorithm, the final orientation of the variational ground state will also be random. If we apply this algorithm two times starting from two different (random) HFB states, one can find that the overlap between the two final HFB states could be $\approx 0$. This subtle dependence in the initial state, although it is irrelevant from the dynamical point of view since the energy is not affected, has notable consequences when analyzing quantum information measures.

\begin{figure*}[t]
    \centering
    \includegraphics[width=1.0\textwidth]{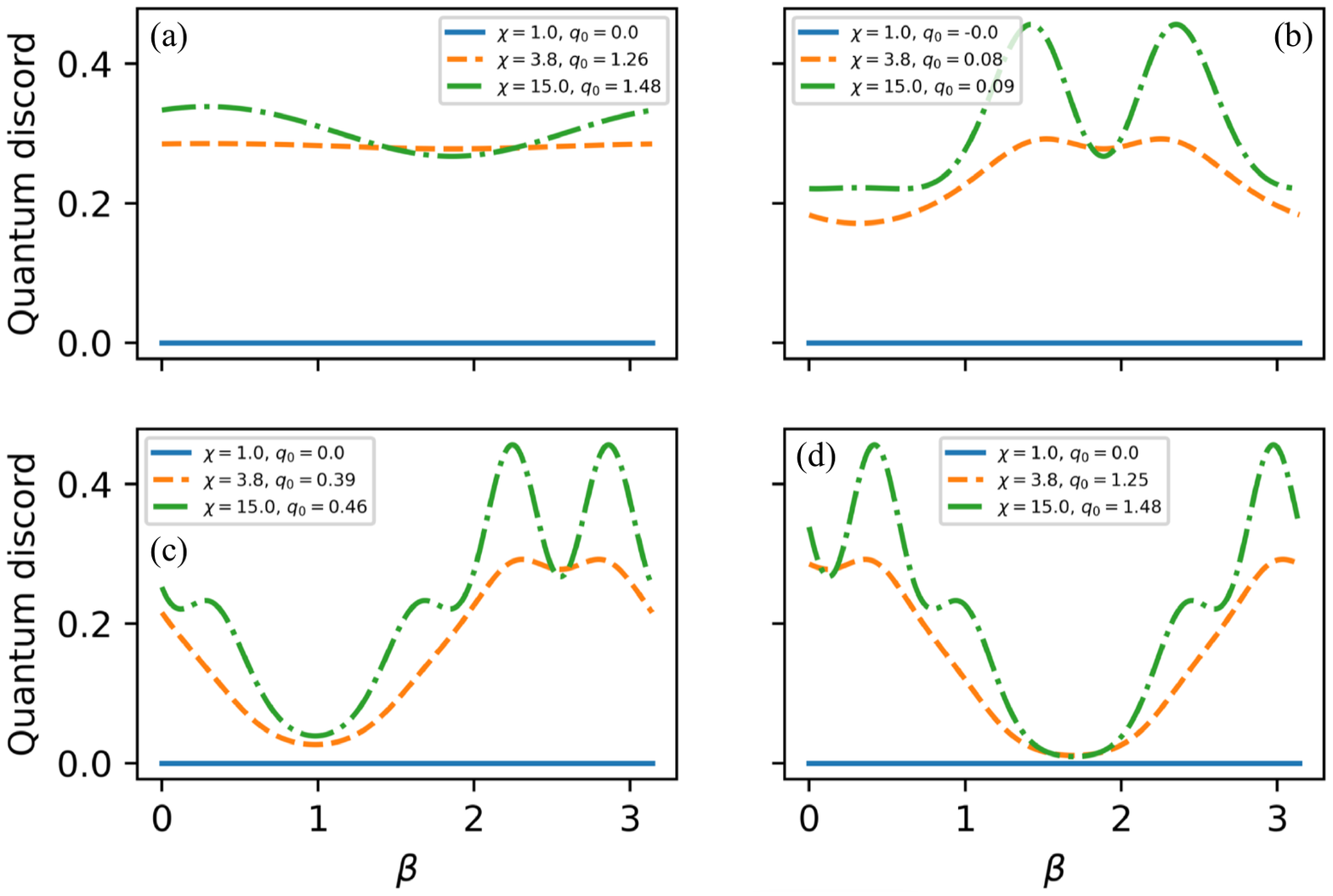}
    \caption{Quantum discord for the HFB ground state computed starting from four different random HFB states. $j=\frac{15}{2}$, $N=10$, $\epsilon=0.2$, $G=0.2$ and $(m,n)=(\frac{1}{2},\frac{3}{2})$. $q_0$ corresponds to the expectation value of the intrinsic quadrupole moment.}
    \label{fig:QD_and_rotations}
\end{figure*}

For the single-$j$ model, the energy of the HFB solution will be invariant under the unitary transformations corresponding to three dimensional rotations, given by the Euler angles $\Omega$, and particle number rotations, given by the gauge angle $\varphi$. For a HFB state defined by the $U$ and $V$ matrices, those transformations correspond to
\begin{equation}
    \begin{split}
        U &\xrightarrow{} D^{(j)}(\Omega)Ue^{i\varphi} \\
        V &\xrightarrow{} D^{(j)}(\Omega)^*Ve^{-i\varphi} \\
    \end{split}
\end{equation}
where the matrix $D^{(j)}(\Omega)$ is the $2j+1$ dimensional representation of the spatial rotation characterized by the Euler angles $\Omega = (\alpha,\beta,\gamma)$. Since the one body density matrix and the pairing tensor can be written in terms of the $U$ and $V$ matrices as $\gamma = V^*V^T$ and $\kappa = V^*U^T$, the symmetry transformations for $\gamma$ and $\kappa$ will be
\begin{equation}
    \begin{split}
        \gamma &\xrightarrow{} D^{(j)}(\Omega)\gamma D^{(j)}(\Omega)^\dagger \\
        \kappa &\xrightarrow{} D^{(j)}(\Omega)\kappa D^{(j)}(\Omega)^T e^{i2\varphi} \\
    \end{split}
\end{equation}
If we now look at the off-diagonal elements of the rotated one body density matrix we observe that they are invariant up to a phase for a $\Omega = (\alpha,0,0)$ and $\varphi$ rotation
\begin{equation}
\begin{split}
    \gamma_{mn} \xrightarrow{} &\sum_{kl}D^{(j)}_{mk}(\Omega)\gamma_{kl} D^{(j)}_{ln}(\Omega)^\dagger \\
    &=\sum_{kl}e^{-im\alpha}\delta_{mk}\gamma_{kl}e^{in\alpha}\delta_{ln} \\
    & = e^{i\alpha(n-m)}\gamma_{mn}
\end{split}
\end{equation}
For the pairing matrix we obtain a similar result
\begin{equation}
    \kappa_{mn} \xrightarrow{} e^{-i\alpha (m+n)}\kappa_{mn}e^{i2\varphi}
\end{equation}
Since the quantum discord between the $(j,m)$ and $(j,n)$ single particle states depends, for a HFB state, on the absolute value of those off diagonal elements and the diagonal ones \cite{faba_1}, it will be invariant. However, if $\Omega = (\alpha,\beta,0)$ with $\beta \in (0,\pi]$, the absolute value of those off diagonal elements won't be invariant and the quantum discord will change. This is illustrated in Fig. \ref{fig:QD_and_rotations} for a $\Omega = (0,\beta,0)$, $\varphi = 0$ type rotation.
There we can observe the behavior of the quantum discord between the $(j,m)$ and $(j,n)$ single particle states for the HFB ground state as a function of the angle $\beta$, where $j = \frac{15}{2}$, $m=\frac{1}{2}$, $n=\frac{3}{2}$, $G = 0.2$, $\epsilon = 0.2$ and $\langle N \rangle = 10$. The blue solid line corresponds to $\chi = 1$ (no deformation), the dashed orange line corresponds to $\chi = 3.8$ (small deformation) and the dash-dotted green line corresponds to $\chi = 15$ (large deformation). Each plot corresponds to a different random initial HFB state.

When the deformation parameter $\chi$ is small enough, there is no deformation (i.e, no rotational symmetry breaking) and the quantum discord is invariant under rotations. Intuitively, the HFB ground state remains spherical, so a rotation doesn't change the orientation. However, when $\chi$ is high enough\footnote{This `high enough' depends on the details of the model such as the Hamiltonian parameters $\epsilon$ and $\chi$, the size of the shell and the expected particle number.} the HFB state goes through a phase transition in which the state is no longer spherical, and therefore it is not invariant under rotations. Since the starting HFB state is random, the final orientation of the ground state will also be random, and the dependence between the quantum discord and the $\beta$ rotation, i.e, the shape of the lines in Fig. \ref{fig:QD_and_rotations}, depends heavily on the initial state, being the dependence more visible when the deformation is high.

At this point, it is important to highlight that the expected value of all observables, such as the energy, the total angular momentum or the particle number do not depend on the single particle basis. They are physically measurable quantities that only depend on the parameter space of the model (in the case of the single-$j$ model, $\epsilon$, $G$ and $\chi$). On the other hand, some quantum information measures such as the quantum discord between single particle states are not observable quantities, and they depend on the single particle basis, which is a mathematical construction used to describe many-body quantum states. Those measures characterize the structure of the quantum state. They don't characterize physical measurable properties. Hence, it is not surprising that those kind of quantities have a high dependency on unitary transformations such as three dimensional or particle number rotations since they can be interpreted as a change of the single particle basis, while observables such as the energy, total angular momentum or particle number remain invariant. However, the eigenvalues of the one body density matrix are, by definition, invariant under unitary transformations. Since its diagonal elements can be interpreted as the occupation numbers of the corresponding single particle basis, this defines a special basis regardless of the symmetry transformations of the Hamiltonian. This is the so called ``natural basis", which is defined as the single particle basis that diagonalizes the one body density matrix. Indeed, the one body entropy, which is a measure of the total correlations of the system, reaches its minimum in the natural orbital basis \cite{gigena_1,gigena_2,tullio_1}. Consequently, analyzing the structure of the quantum state through correlations between elements of the natural basis offers two key advantages: first, the dependency between the quantum information quantities and arbitrary symmetry transformations is removed; second, since the natural orbital basis minimizes the total correlations, a quantum information analysis in this basis guarantees that only essential correlations are taken into account. Hence, all the results presented in this work correspond with an analysis in the natural orbital basis.

\subsection{\label{sec:results_corr_structure_HFB}Correlation structure of the HFB ground state}

The main characteristic of the HFB solution for the ground state of the single-$j$ Hamiltonian is the symmetry breaking of the rotational and particle number symmetries \cite{dobaczewski_1,Baranger1965}. While the particle number breaking is more or less evident because of the nature of the HFB ansatz, the rotational symmetry breaking comes from the interplay between the pairing force (whose intensity is given by the parameter $G$) that favors the system to be in a spherical shape, and the quadrupole force (whose intensity is given by the parameter $\chi$) that forces the system to deform. This "tug of war" between those two forces causes a quantum phase transition from a spherical phase to a deformed phase that can be studied through quantum information measures.

\begin{figure}[h!]
    \centering
    \includegraphics[width=0.48\textwidth]{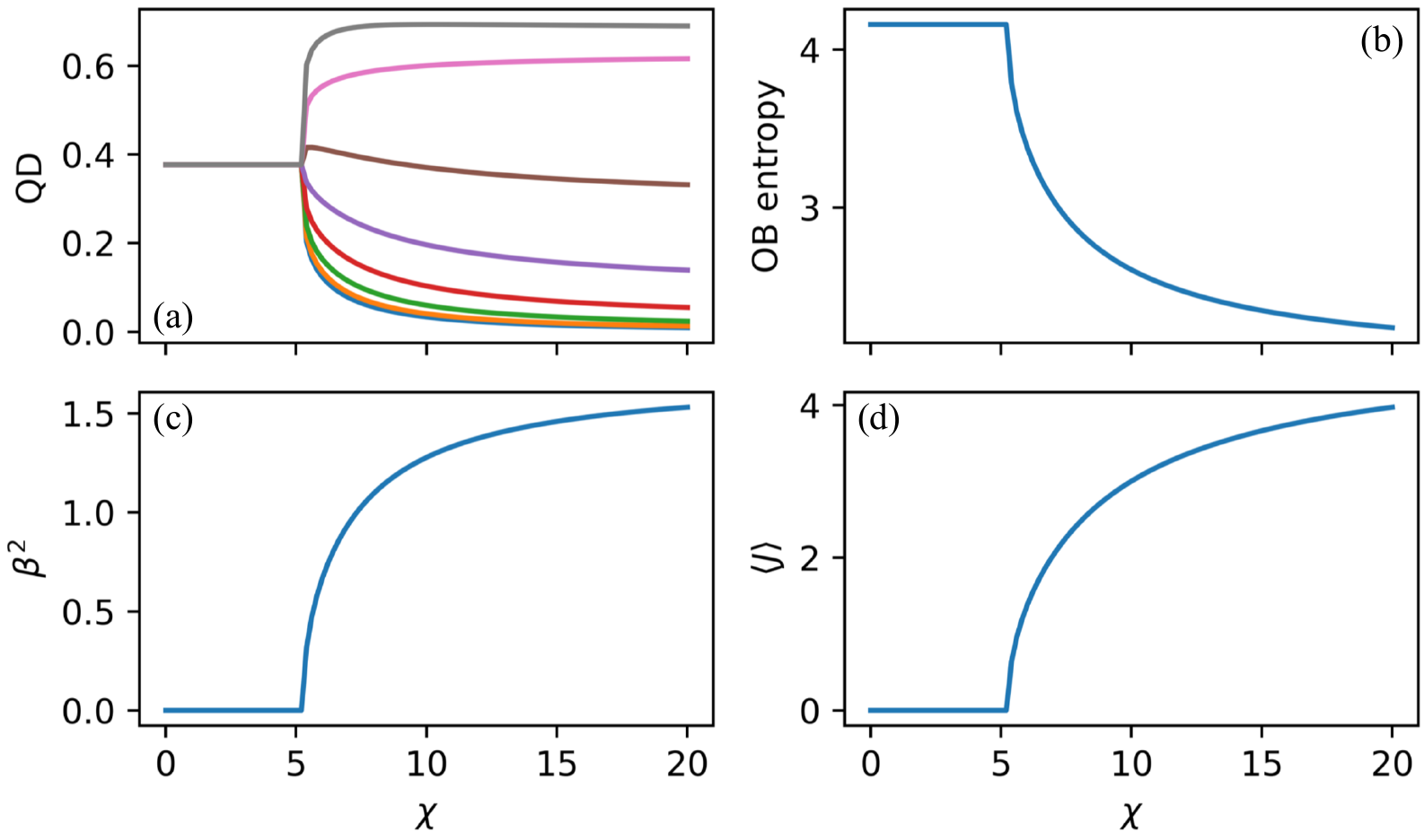}
    \caption{(a) QD, (b) one body entropy, (c) quadrupole deformation and (d) total angular momentum for the HFB ground state in the single-$j$ model with $j = \frac{15}{2}$ and $N=2$, as a function of the deformation parameter $\chi$. $\epsilon=0.2$ and $G=0.2$. Each line of panel (a) of Figs. \ref{fig:MFj15N2}, \ref{fig:MFj15N4}, \ref{fig:MFj15N6} and \ref{fig:MFj15N8} represents QD between two paired orbitals, this is, $(j,m)$ and $(j,-m)$ states. For readability, the specific $m$ values corresponding to each line are not labeled, as they are not relevant to the discussion.}
    \label{fig:MFj15N2}
\end{figure}

\begin{figure}[h!]
    \centering
    \includegraphics[width=0.48\textwidth]{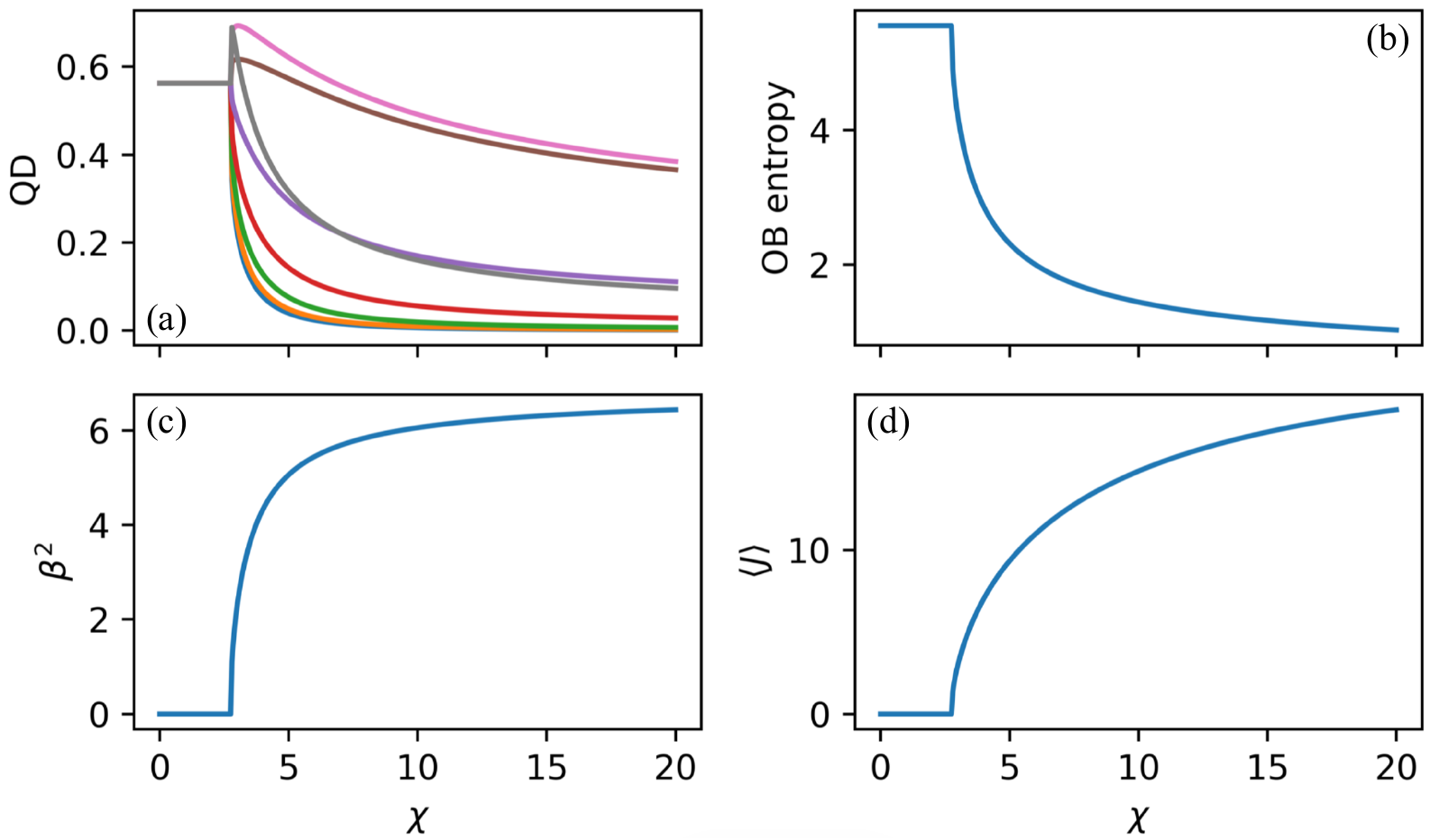}
    \caption{Same as Fig. \ref{fig:MFj15N2} but for $N=4$.}
    \label{fig:MFj15N4}
\end{figure}

\begin{figure}[h!]
    \centering
    \includegraphics[width=0.48\textwidth]{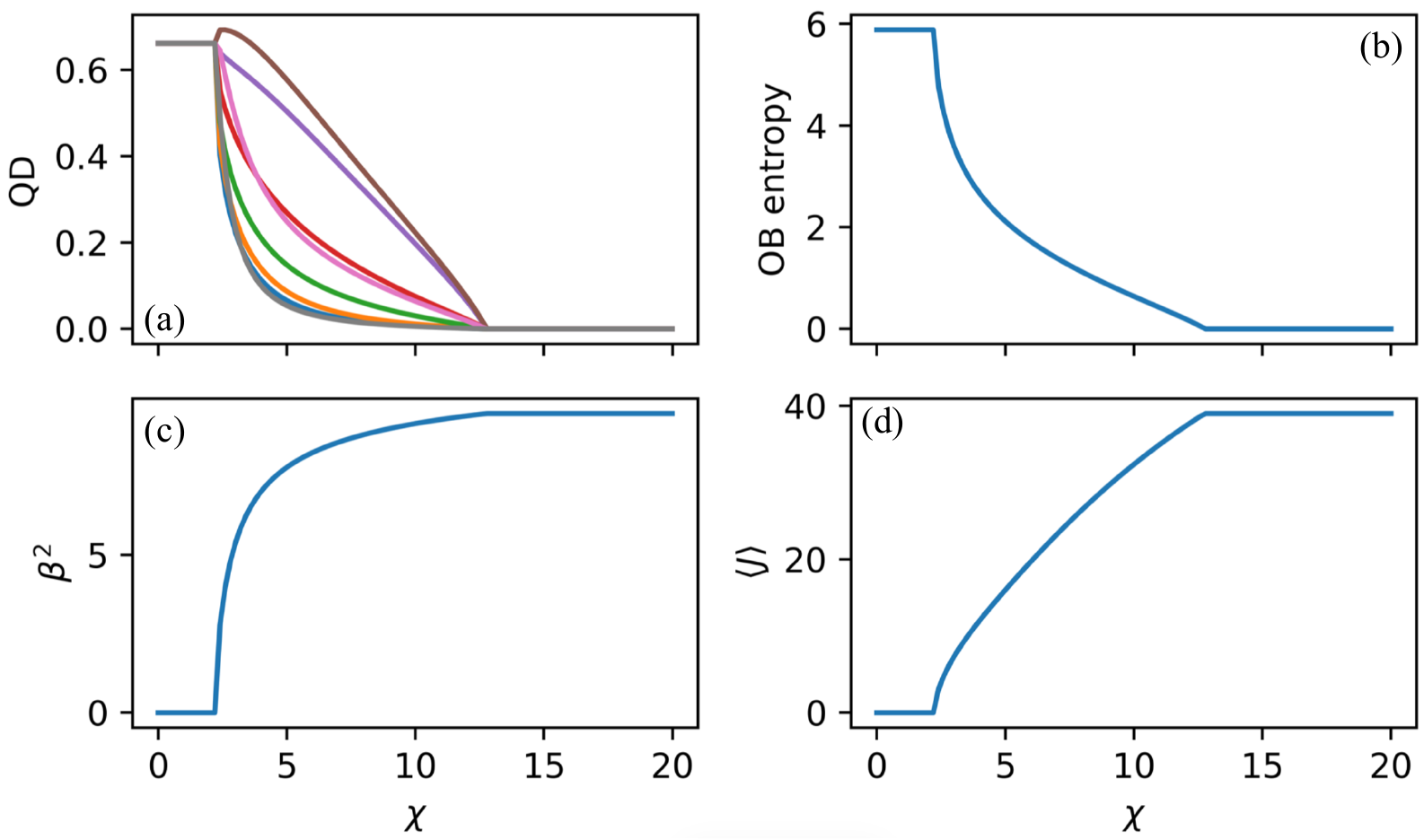}
    \caption{Same as Fig. \ref{fig:MFj15N2} but for $N=6$.}
    \label{fig:MFj15N6}
\end{figure}

\begin{figure}[h!]
    \centering
    \includegraphics[width=0.48\textwidth]{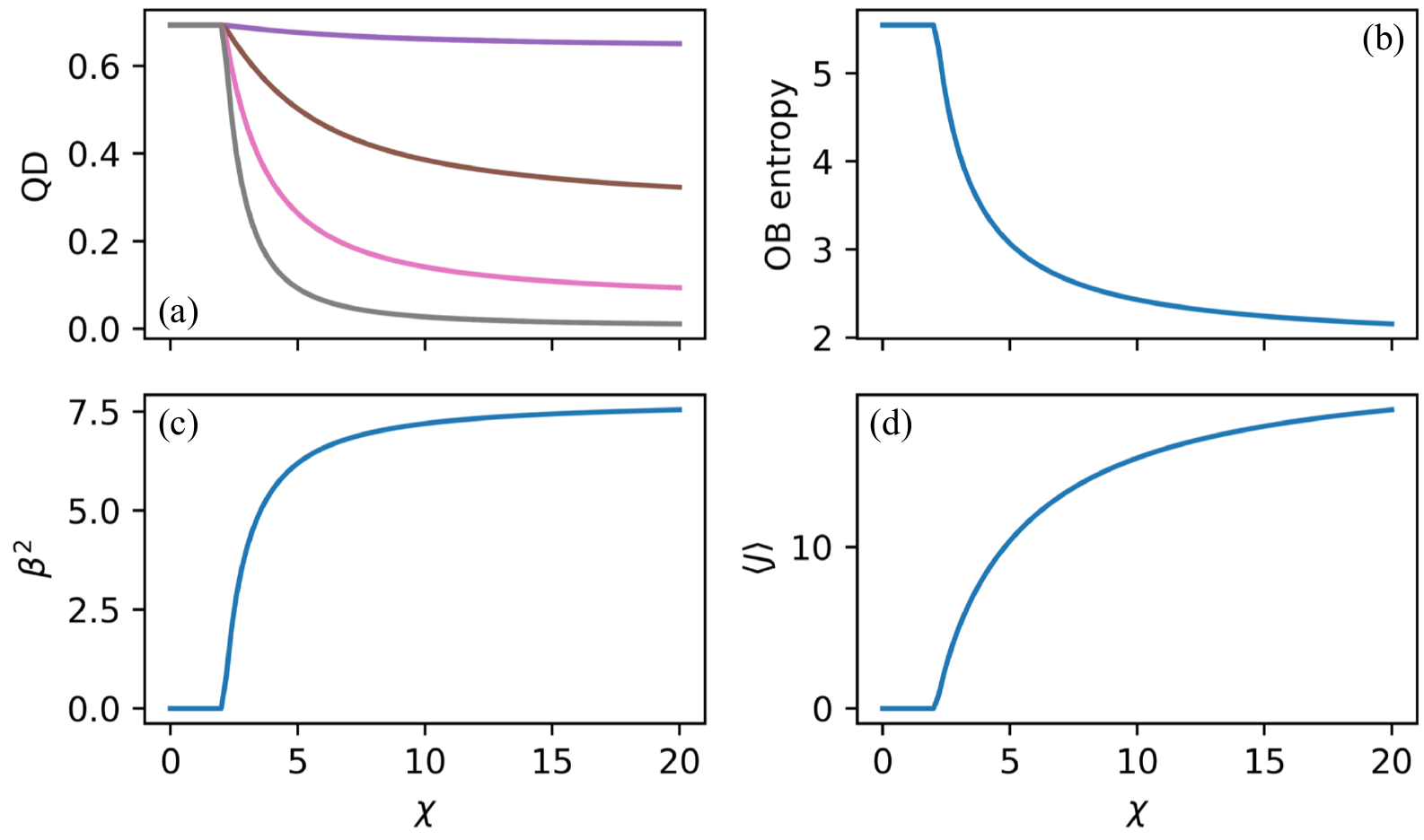}
    \caption{Same as Fig. \ref{fig:MFj15N2} but for half-filling.}
    \label{fig:MFj15N8}
\end{figure}

Specifically, we can use the quantum discord and mutual information as a measure of quantum and total correlations, respectively, between single particle states. Since we are interested in the study of intrinsic correlations, we work in the natural orbital basis. Hence, quantum correlations are generated exclusively by pairing \cite{faba_1}. Because of the canonical form of the pairing matrix, natural orbitals are correlated by pairs, and it can be shown that the reduced quantum state that is obtained by tracing out all single particle states but two paired ones is pure (see Appendix \ref{sec:AppendixA}). Taking into account that the total state is pure, there are no quantum nor classical correlations between one pair of correlated orbitals and other. Moreover, quantum and classical correlations between two paired single particle states are exactly the same and it's value is the von Neumann entropy of the reduced state \cite{two_qubit}. 
Hence, it is enough to compute quantum discord between two paired orbitals to obtain quantum, classical and total correlations. In Figs. \ref{fig:MFj15N2}, \ref{fig:MFj15N4}, \ref{fig:MFj15N6} and \ref{fig:MFj15N8} the quantum discord, together with the one body entropy, the expected value of the norm of the quadrupole operator and the total angular momentum is shown as a function of the deformation parameter $\chi$ and for different values of the expected particle number and the size ($j$) of the shell.
The one body entropy, deformation parameter\footnote{Be aware that we are calling "deformation parameter" two different quantities. The first one is the $\chi$ factor that quantifies the intensity of the quadrupole term of the hamiltonian. The second one is $\beta^2$, which is defined as $\beta^2 = \sum_\mu \abs{\langle Q_\mu\rangle }^2$, and it is a measure of the deformation of the quantum state.} $\beta^2$ and total angular momentum have essentially the same behavior for $N$ = 2, 4, and 8 (Figs. \ref{fig:MFj15N2}, \ref{fig:MFj15N4} and \ref{fig:MFj15N8} respectively). For values of $\chi$ with $\chi\leq\chi_c$ (where $\chi_c$ is the critical point and its value depends on the Hamiltonian parameters), $\langle J^2\rangle = \beta^2 =  0$, which means that there is no deformation and the system is in a spherical phase. The angular momentum symmetry is not broken: since $\langle J^2\rangle$ is a non-negative quantity and $\langle J^2\rangle =  0$, there is no quantum fluctuations in the total angular momentum and the HFB ground state has a well defined zero angular momentum. This means that all diagonal elements of the one body density matrix acquire the same value, $\frac{N}{2j+1}$ (see Appendix \ref{sec:AppendixB}) \cite{faba_thesis}, which is independent of $\chi$. This behavior changes when $\chi \geq \chi_c$. At this point ($\chi_c$), the system overcomes a QPT to a deformed phase since $\beta^2 \neq 0$. In this phase, the total angular momentum is no longer a well defined quantum number ($\langle J^2 \rangle \neq 0$) and this spontaneous symmetry breaking is reflected in the one body entropy, which is no longer constant. In fact, total correlations decrease with $\chi$, and the quantum/classical correlations are redistributed. This redistribution can be seen through the QD between two paired orbitals, and depends on $j$ and $N$. We will discuss this later.

As just discussed, the $N$ = 2, 4, and 8 cases (Figs. \ref{fig:MFj15N2}, \ref{fig:MFj15N4} and \ref{fig:MFj15N8}) are very similar. However, when $N $ = 6 (Fig. \ref{fig:MFj15N6}) one can observe another critical point $\chi_c'$ with $\chi_c'>\chi_c$. When $\chi \geq \chi_c'$ the total correlations are zero, and both the total angular momentum and the deformation parameter remain constant. Since the one body entropy is zero, its eigenvalues must be 0 or 1. Hence, the state is a Slater determinant, which means that the particle number symmetry is ``spontaneously restored". Although this sounds counter-intuitive in a first sight, it can be explained if we take into account the intensity of the pairing force and the expected value of the angular momentum. The pairing force is quantified by the parameter $G$, which remains constant ($G = 0.2$). When $\chi$ increases, the relative strength of the pairing term in the Hamiltonian with respect to the quadrupole force decreases. If we relate the particle number symmetry breaking to the pairing strength, then it is reasonable to state that if $G$ is low enough with respect to $\chi$, the pairing force is not strong enough to break the particle number symmetry. Hence, the point $\chi_c'$ is a critical point in the parameter space where the particle number symmetry breaks if we read the figure `backwards' (from $\chi > \chi_c'$ to $\chi=\chi_c'$, this is, when $\frac{G}{\chi}$ is increasing). However, this symmetry breaking process depends also on $j$ and $N$. For the $j = \frac{15}{2}$ shell, if $N\neq 6$, the critical point $\chi_c'$ has not been observed (Figs. \ref{fig:MFj15N2}, \ref{fig:MFj15N4} and \ref{fig:MFj15N8}): since the quantum discord is non zero, the pairing tensor is non zero and the particle number symmetry is always broken. For the $N = 6$ case, this critical point $\chi_c'$ is observed because, for this specific configuration, significantly higher angular momentum values are reached compared to the other cases. This high angular momentum regime is needed to totally break pairing correlations, thereby restoring particle number symmetry. However, a complete explanation of this mechanism lies beyond the scope of the present study and would require a dedicated investigation.

Although the behavior of the total correlations (one body entropy), $\beta^2$ and $\langle J^2\rangle$ when $N$ = 2, 4, and 8 is essentially the same, it is not the case for the quantum correlations. If $\chi < \chi_c$, quantum discord remains constant with the same value for all pairs of paired natural orbitals. This property, which is common for all $N$, comes from the fact that all natural occupation numbers are $\frac{N}{2j+1}$ because of the zero angular momentum. Hence, quantities involved in the quantum discord (the one body density matrix and pairing tensor) do not depend on any parameter but $j$ and $N$. This argument is valid since the state is of mean field type.

When $\chi>\chi_c$, the behavior of the quantum discord strongly depends on $N$. When $N$ = 2 (Fig. \ref{fig:MFj15N2}, (a)), only two pairs of natural orbitals became more correlated as the deformation increases, while the rest show a decreasing behavior which can be subtle (brown line, in the middle) or very pronounced (lower lines). For $N$ =4, the correlation between all pairs decreases with $\chi$, although some of them show a peak near $\chi_c$. However, a different behavior occurs due to the particle number breaking in $\chi_c'$ for $N$ =6. The state is a Slater determinant for the region $\chi>\chi_c'$, and correlations between two paired orbitals are zero. Hence, a second abrupt change in the behavior of QD (apart from the one in $\chi_c$) occurs in $\chi_c'$. Here we can see the the first difference between the particle number symmetry and the $J$ = 0 symmetry: while the particle number symmetry ($\chi>\chi_c'$ region) kills all correlations between two orbitals, the $J=0$ symmetry ($\chi<\chi_c$ region) maximizes them. This can be understood taking into account that the HFB ground state is composed of uncorrelated Cooper pairs\footnote{Typically Cooper pairs are associated with a subsystem formed by two fermions with opposite spin an linear momentum. This idea can be extended to a subsystem formed by one orbital and its corresponding time reversed one, so from now on we will call this concept Cooper pair. Note that a Cooper pair defined this way doesn't have necessarily a zero angular momentum, since the orbitals doesn't have necessarily a well defined angular momentum. This extension of the Cooper pair concept is typical in the nuclear physics literature \cite{RingSchuck1980}.}, since all reduced states composed of two paired natural orbitals are always pure (see Appendix \ref{sec:AppendixA}). For the spherical phase ($\chi<\chi_c$), all Cooper pairs have zero angular momentum, so paired natural orbitals are highly correlated. However, for the deformed phase ($\chi>\chi_c$) Cooper pairs must have a non zero angular momentum or a non defined angular momentum\footnote{Since natural single particle states don't necessarily have a well defined single particle angular momentum, non zero angular momentum Cooper pairs don't have necessarily a well defined angular momentum.}. In any case, those Cooper pairs contribute to the expected value of the total angular momentum while they remain uncorrelated between them. As the deformation $\chi$ increases, the contribution of each Cooper pair to the total angular momentum grows, which is translated into less correlation between the two paired orbitals that define the corresponding Cooper pair. Finally, when $\chi>\chi_c'$, all natural orbitals are completely uncorrelated (the Cooper pair breaks), the maximum angular momentum is reached, the pairing tensor becomes zero and the particle number symmetry is restored. It is worth noting that the previously described coupling scheme for the spherical phase (this is, when the total angular momentum comes from the individual contribution of the angular momentum of each Cooper pair) is not the only one. As we will see in the next section, $J=0$ projection for the symmetry broken phase increases the complexity of the coupling scheme, adding correlation between Cooper pairs. 

At half filling ($N = 8$), QD exhibits a distinct behavior: it is pairwise degenerate because each conjugate pair of natural orbitals shares identical correlations with a corresponding partner pair for all $\chi$. This can be explained through the particle-hole symmetry. Paired orbitals are related to other paired orbitals through a particle hole transformation, this is, $N \xrightarrow{} 2j+1-N$. In the half filling case ($N=8$ for $j=\frac{15}{2}$) this transformation leaves the system invariant since $N \xrightarrow{} 2j+1-N = N$ Hence, correlations between a paired orbital and its corresponding particle-hole transformed pair must be exactly the same.

As a final remark for this section it is worth to note that, since the reduced state corresponding to two paired natural orbitals is pure for all pairs, the mean-field ground state can be though as a product state of orbital pairs. This is a consequence of the mean-field ansatz, so it holds for all Hamiltonians. Hence, the only possible source of correlations for this state are, by construction, the correlations within each pair. Taking into account that, for the same reason, the only contributions for the total angular momentum of the system come from the angular momentum of each pair, we can establish a link between the correlation structure and the total angular momentum (i.e., the rotational symmetry). Following this idea, we observe that for the spherical region, all pairs are maximally correlated (on average) and the angular momentum is zero. However, when the rotational symmetry is broken, the correlations within each pair decrease (on average), thereby reducing the total correlations. This is opposed to the generally accepted idea in which the system breaks the symmetry to incorporate more correlations. In this case could be a consequence of the interaction between the rotational symmetry and the particle number symmetry since, as we have observed, the restoration of this last symmetry reduces the correlation within each pair to zero.

\subsection{\label{sec:results_corr_structure_J0}Correlation structure of the HFB ground state with $J=0$ restoration}

\begin{figure}[h!]
    \centering
    \includegraphics[width=0.48\textwidth]{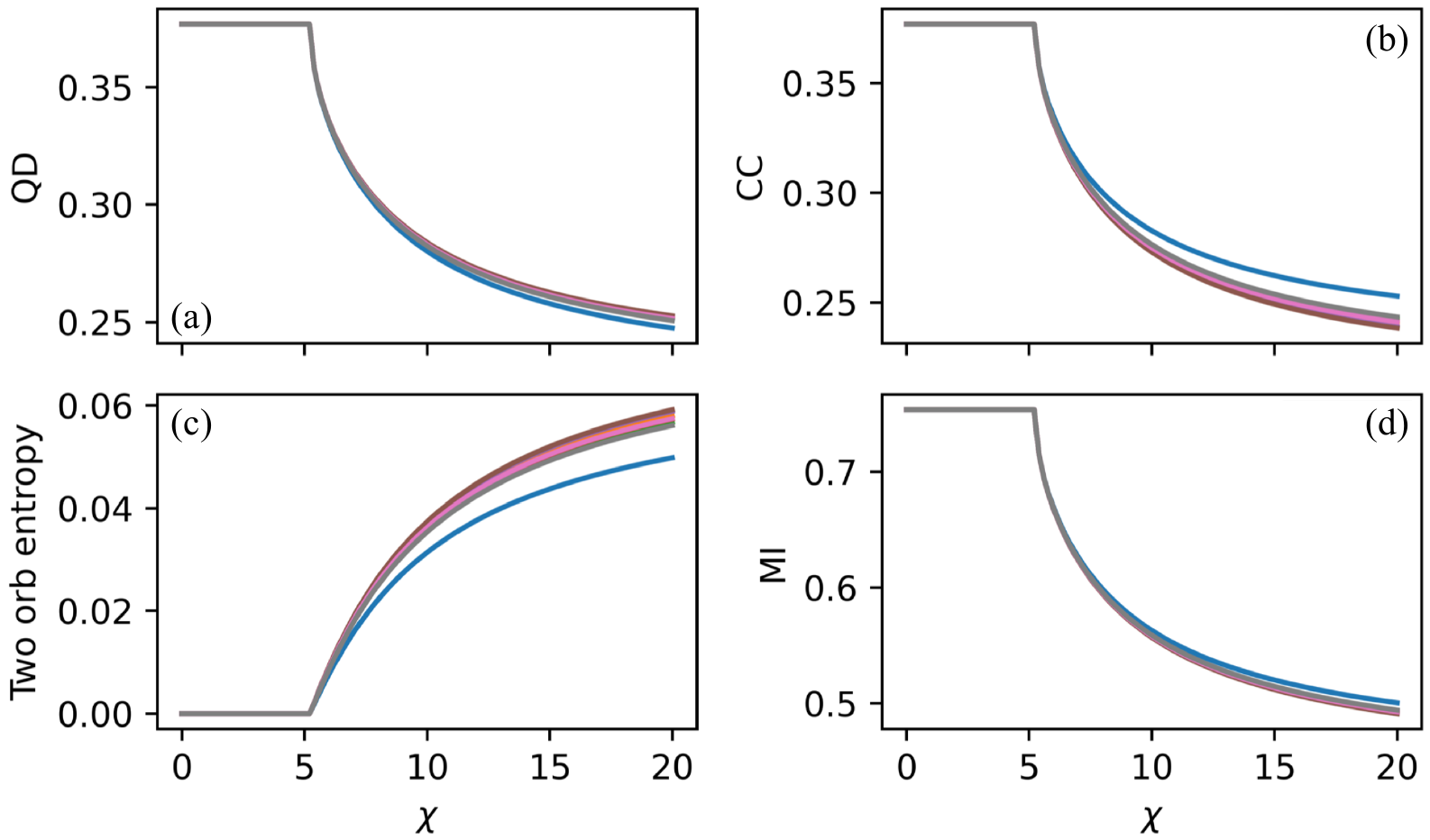}
    \caption{(a) QD, (b) classical correlations, (c) two orbital entropy and (d) mutual information for the HFB ground state with $J=0$ restoration in the single-$j$ model with $j = \frac{15}{2}$ and $N=2$. $\epsilon=0.2$ and $G=0.2$.}
    \label{fig:MFJ0j15N2}
\end{figure}

\begin{figure}[h!]
    \centering
    \includegraphics[width=0.48\textwidth]{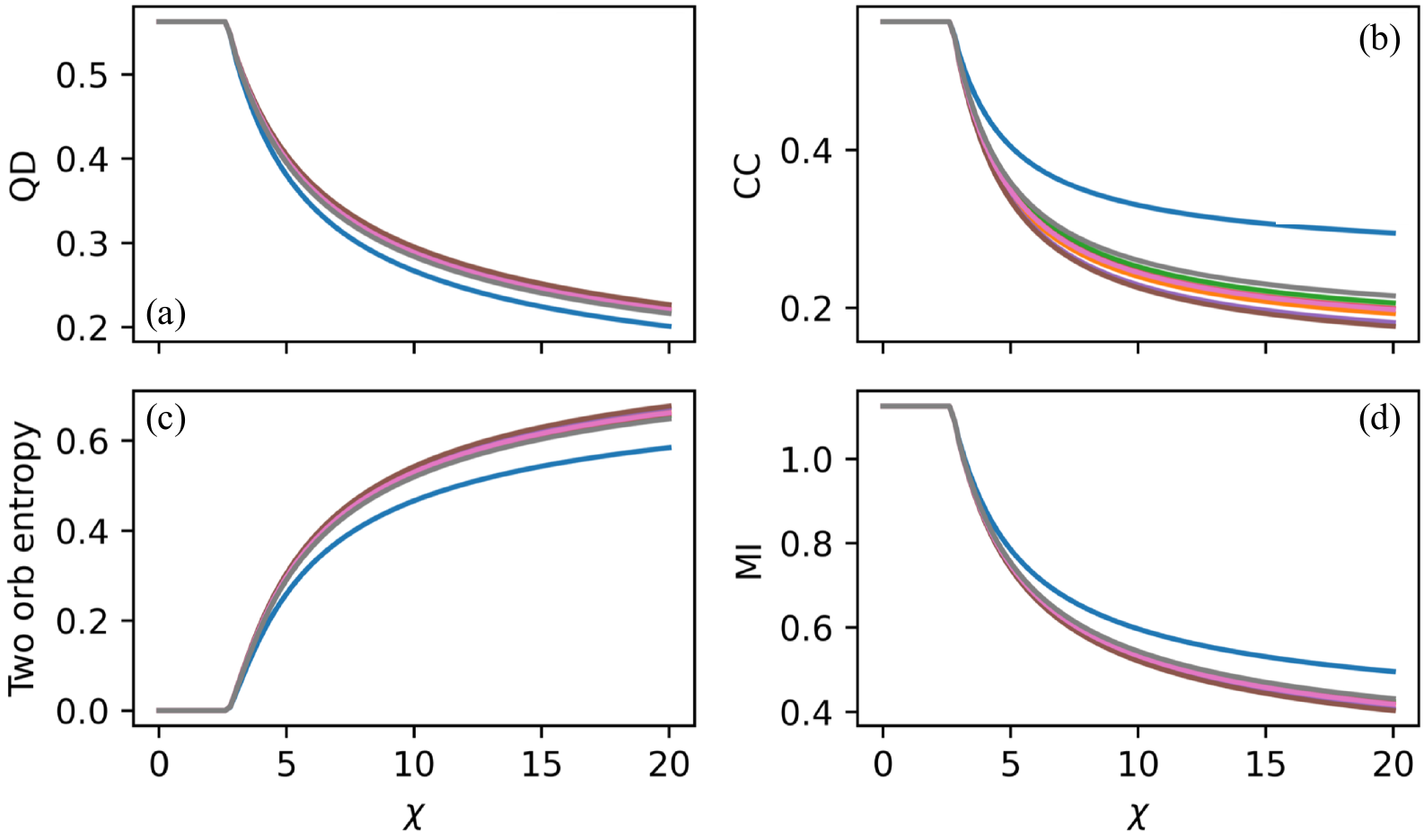}
    \caption{Same as Fig. \ref{fig:MFJ0j15N2} but for $N=4$.}
    \label{fig:MFJ0j15N4}
\end{figure}

\begin{figure}[h!]
    \centering
    \includegraphics[width=0.48\textwidth]{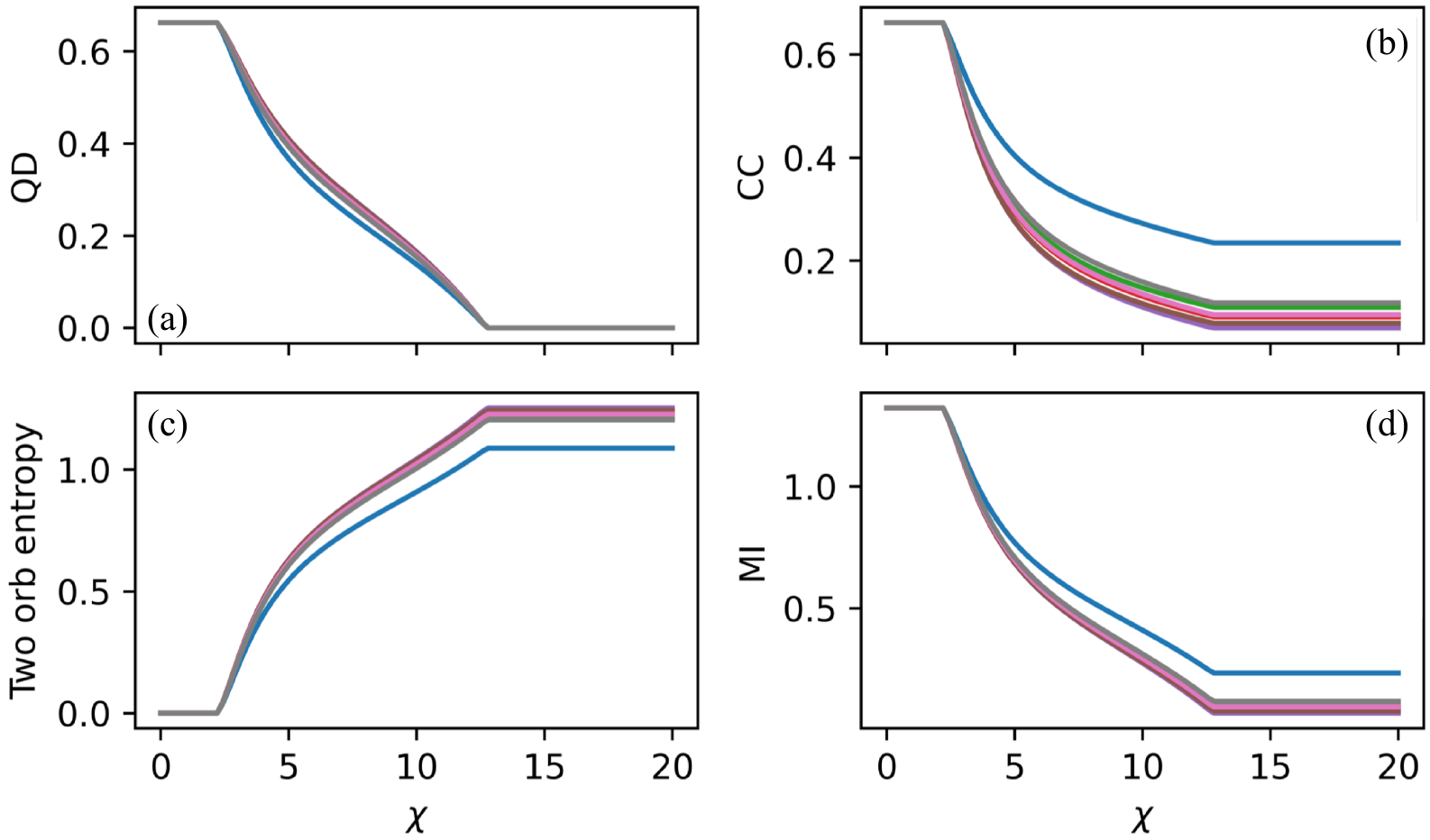}
    \caption{Same as Fig. \ref{fig:MFJ0j15N2} but for $N=6$.}
    \label{fig:MFJ0j15N6}
\end{figure}

\begin{figure}[h!]
    \centering
    \includegraphics[width=0.48\textwidth]{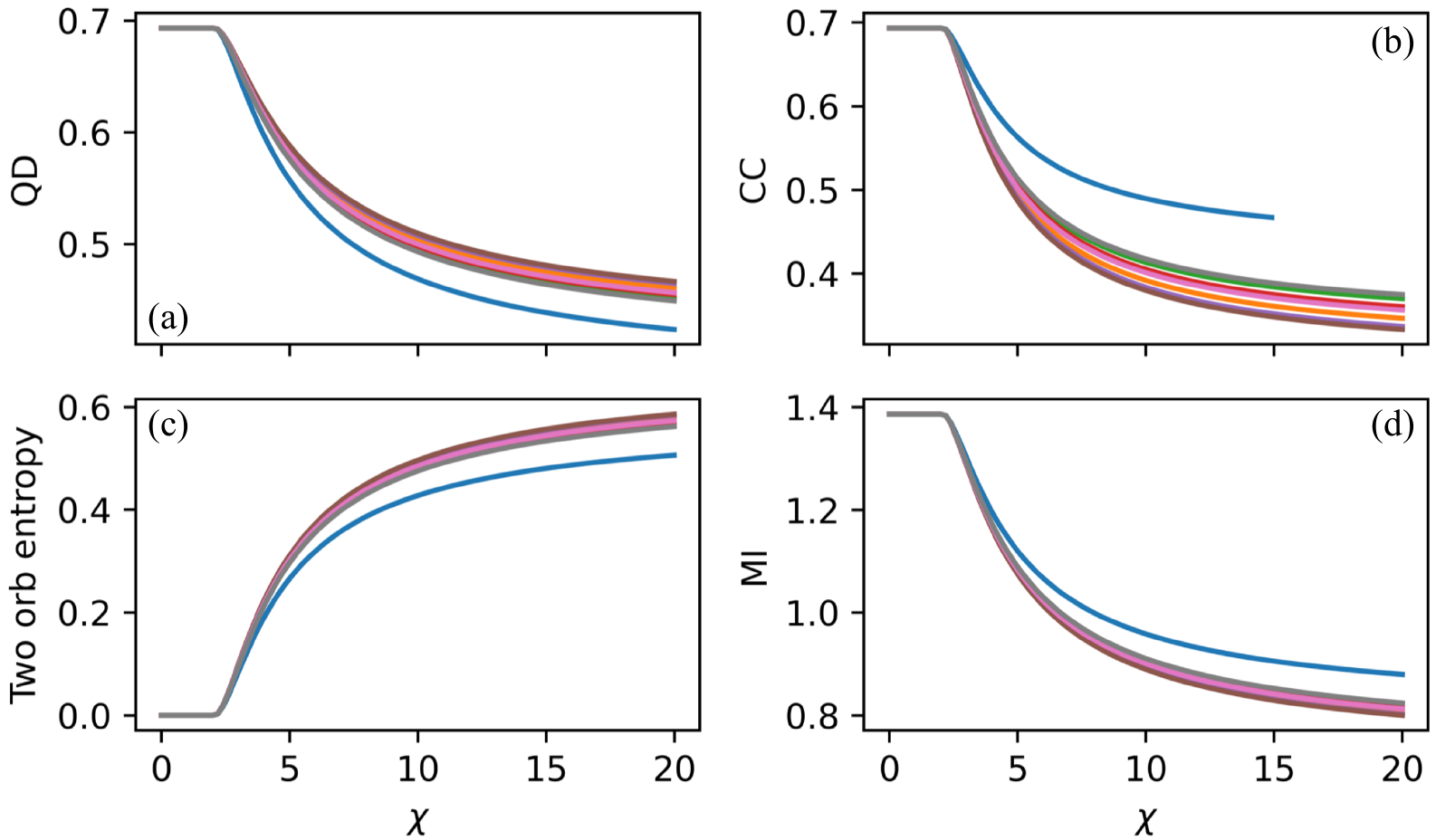}
    \caption{Same as Fig. \ref{fig:MFJ0j15N2} but for half-filling.}
    \label{fig:MFJ0j15N8}
\end{figure}

As discussed in Sec. \ref{sec:symmetry_restoration}, the natural orbital basis for a quantum state with well defined angular momentum $J=0$ is simply the initial single particle states $(j,m)$, so intrinsic correlations can be analyzed through this basis. In order to compare with the previous results, we will show figures for the same $j = \frac{15}{2}$ shell and particle number $N$ = 2, 4, 6, and 8 (Figs. \ref{fig:MFJ0j15N2}, \ref{fig:MFJ0j15N4}, \ref{fig:MFJ0j15N6} and \ref{fig:MFJ0j15N8} respectively). 


The $J=0$ restored state is not a product state in general, meaning that the two orbital reduced density matrix for paired single particle natural states, $(j,m)$ and $(j,-m)$, is not pure in the general case. Hence, correlations between this subsystem (a Cooper pair) and the rest may exist, and can be measured through the two orbital entropy, since the total quantum state is still pure. 
Quantum discord between $(j,m)$ and $(j,-m)$ doesn't have to be equal to classical correlations, and the von Neumann entropy of the reduced state formed by the single particle state $(j,m)$ is no longer valid to measure quantum correlations between $(j,m)$ and $(j,-m)$. However, it is possible to compute these quantum correlations using the general expression in \cite{faba_1}.

Figs. \ref{fig:MFJ0j15N2}, \ref{fig:MFJ0j15N4} \ref{fig:MFJ0j15N6} and \ref{fig:MFJ0j15N8} reveal that, in the spherical phase, the symmetry restoration has no effect on the quantum information quantities: QD remains the same as in the HFB state and the two orbital entropy is zero. Differences arise for the deformed phase ($\chi >\chi_c$). Now the two orbital reduced state ($(j,m)$ and $(j,-m)$) is not pure anymore, and correlation between this subsystem and the rest of the system grows with the quadrupole force parameter $\chi$. Under the Cooper pair picture, this can be associated to interactions between Cooper pairs due to the more complex $J=0$ coupling scheme that arises as a consequence of the symmetry restoration applied to the deformed HFB state. On the other hand, QD, classical correlations and mutual information within Cooper pairs decrease with $\chi$. Since the one body entropy remains constant, this indicates that correlations within Cooper pairs are transformed into correlations among Cooper pairs as a consequence of the angular momentum symmetry restoration.

On the other hand, interesting properties can be seen for the $N$ = 6 case (Fig. \ref{fig:MFJ0j15N6}). QD follows a more uniform behavior for all $(j,m)$ and $(j,-m)$ pairs than in the HFB case, and for $\chi>\chi_c'$ quantum correlations are zero in both cases. However, this is not true for classical correlations: for the symmetry restored state, they remain constant but non-zero when $\chi>\chi_c'$. The effect of the symmetry restoration for the well defined particle number region ($\chi>\chi_c'$) is to incorporate purely classical correlations within the Cooper pairs. Since the single particle basis $(j,m)$ is always the natural one (as discussed in Sec. \ref{sec:symmetry_restoration}), and the $J=0$ projection doesn't break the particle number, it is impossible to have $\gamma_{m,-m} \neq 0$ or $\kappa_{m,-m} \neq 0$ for the $\chi>\chi_c'$ region, forcing the quantum correlations to be always zero.

\subsection{\label{sec:results_corr_structure_NJ0}Correlation structure of the HFB ground state with particle number and $J=0$ restoration}

When restoring both particle number and angular momentum $J=0$ quantum numbers, the natural orbital basis will be formed by the single particle states $(j,m)$ because of the angular momentum symmetry, and the pairing tensor will be zero because of the well defined particle number. Hence, QD between two orbitals will be zero, and mutual information will be a measure of classical correlations. On the other hand, since the product state structure is broken (in general), correlations among Cooper pairs will be non zero (in general). Again, following the previous discussions, this is a consequence of the more complex coupling scheme for the total angular momentum, which can be interpreted in terms of Cooper pairs as an interaction among them. In the following figures (Figs. \ref{fig:MFJ0Nj15N2}, \ref{fig:MFJ0Nj15N4}, \ref{fig:MFJ0Nj15N6} and \ref{fig:MFJ0Nj15N8}), only the two orbital entropy and the mutual information between two paired orbitals is shown, since they are the more relevant quantum information quantities in order to compare with the previous results.
\begin{figure}[h!]
    \centering
    \includegraphics[width=0.48\textwidth]{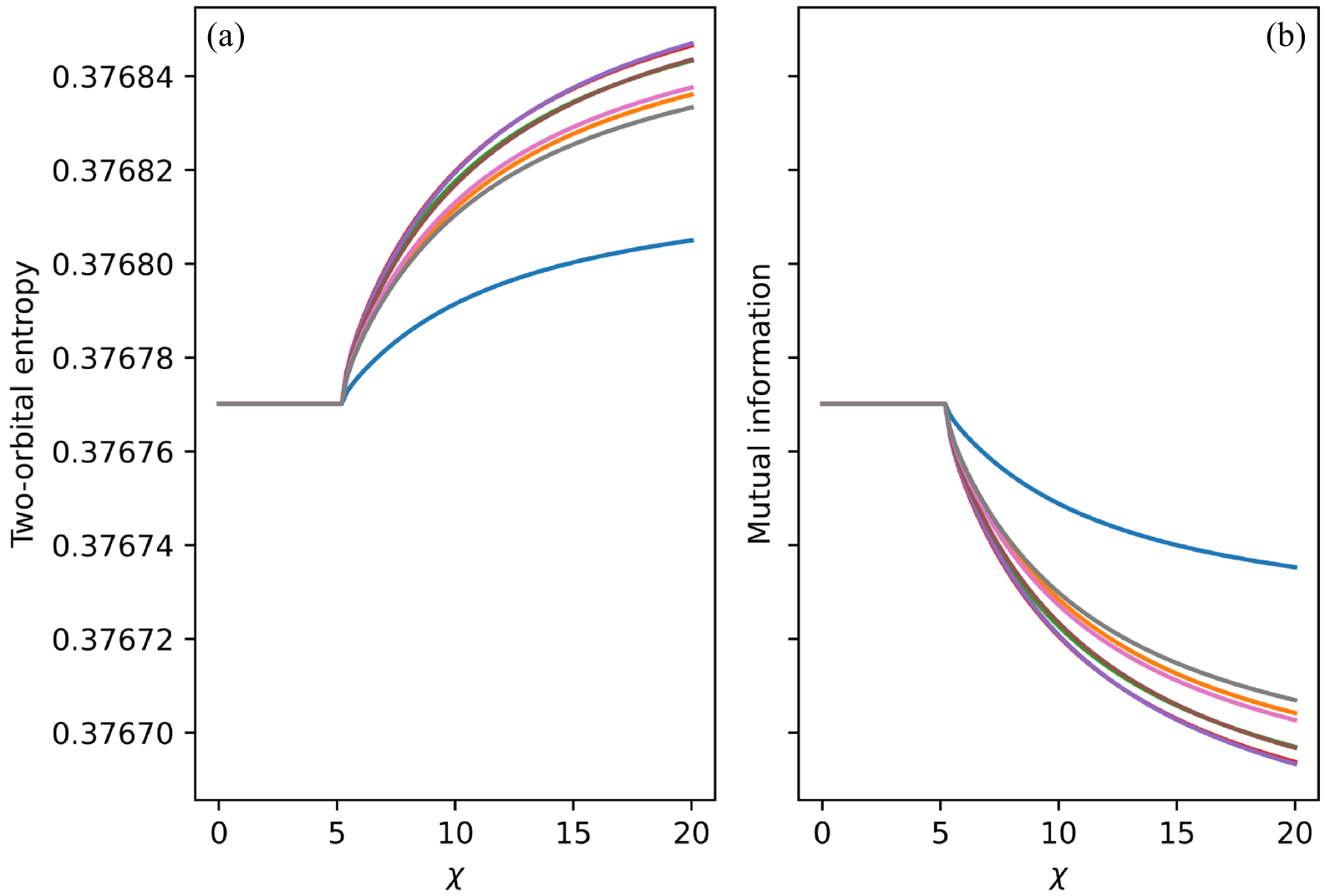}
    \caption{(a) Two orbital entropy for the reduced state $\rho^{(m,-m)}$ and (b) mutual information between the $(j,m)$ and $(j,-m)$ orbitals for the fully restored state (see main text), with $N = 2$ and $j = \frac{15}{2}$. $\epsilon=0.2$ and $G=0.2$.}
    \label{fig:MFJ0Nj15N2}
\end{figure}
\begin{figure}[h!]
    \centering
    \includegraphics[width=0.48\textwidth]{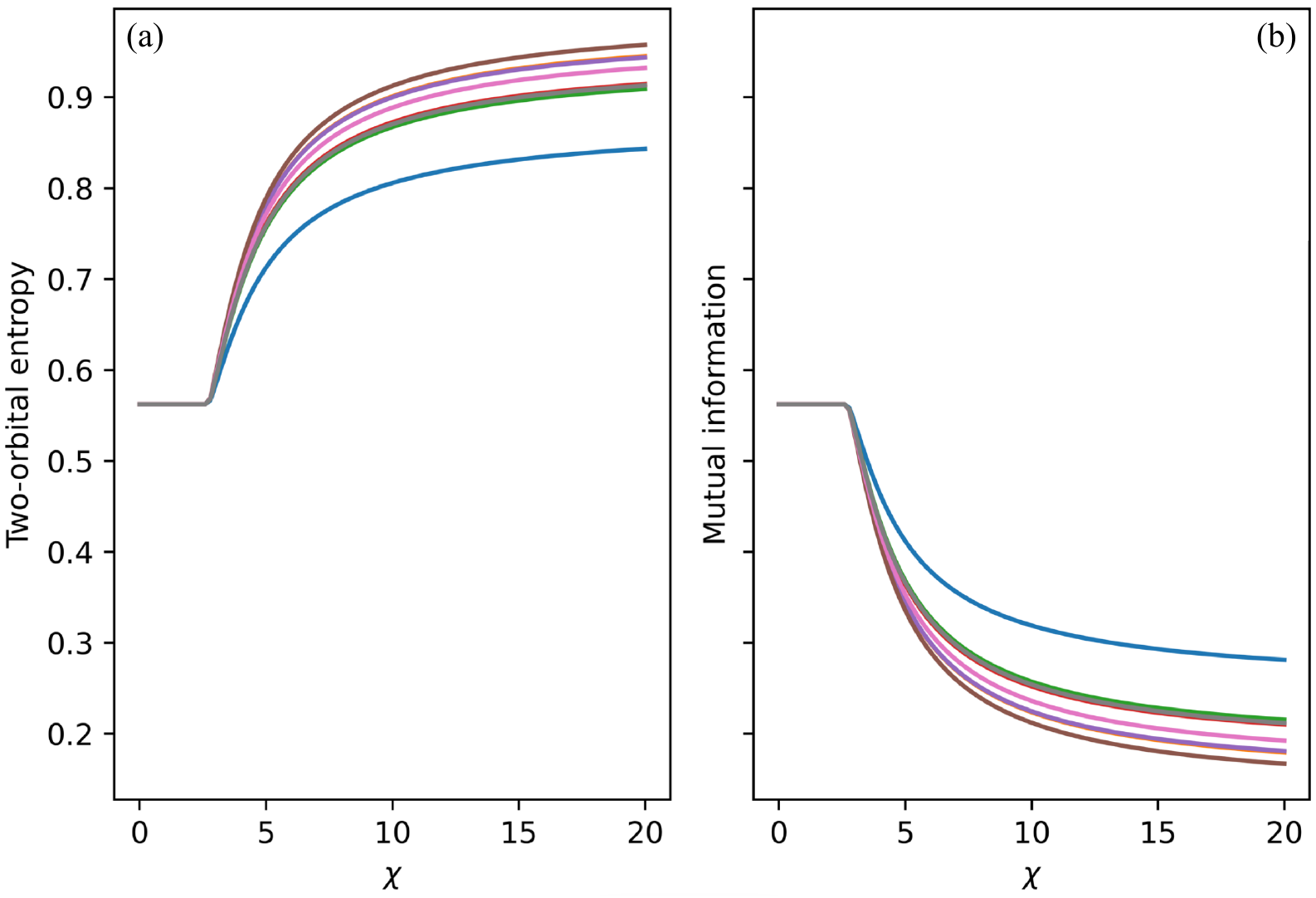}
    \caption{Same as Fig. \ref{fig:MFJ0Nj15N2} but for $N=4$.}
    \label{fig:MFJ0Nj15N4}
\end{figure}
\begin{figure}[h!]
    \centering
    \includegraphics[width=0.48\textwidth]{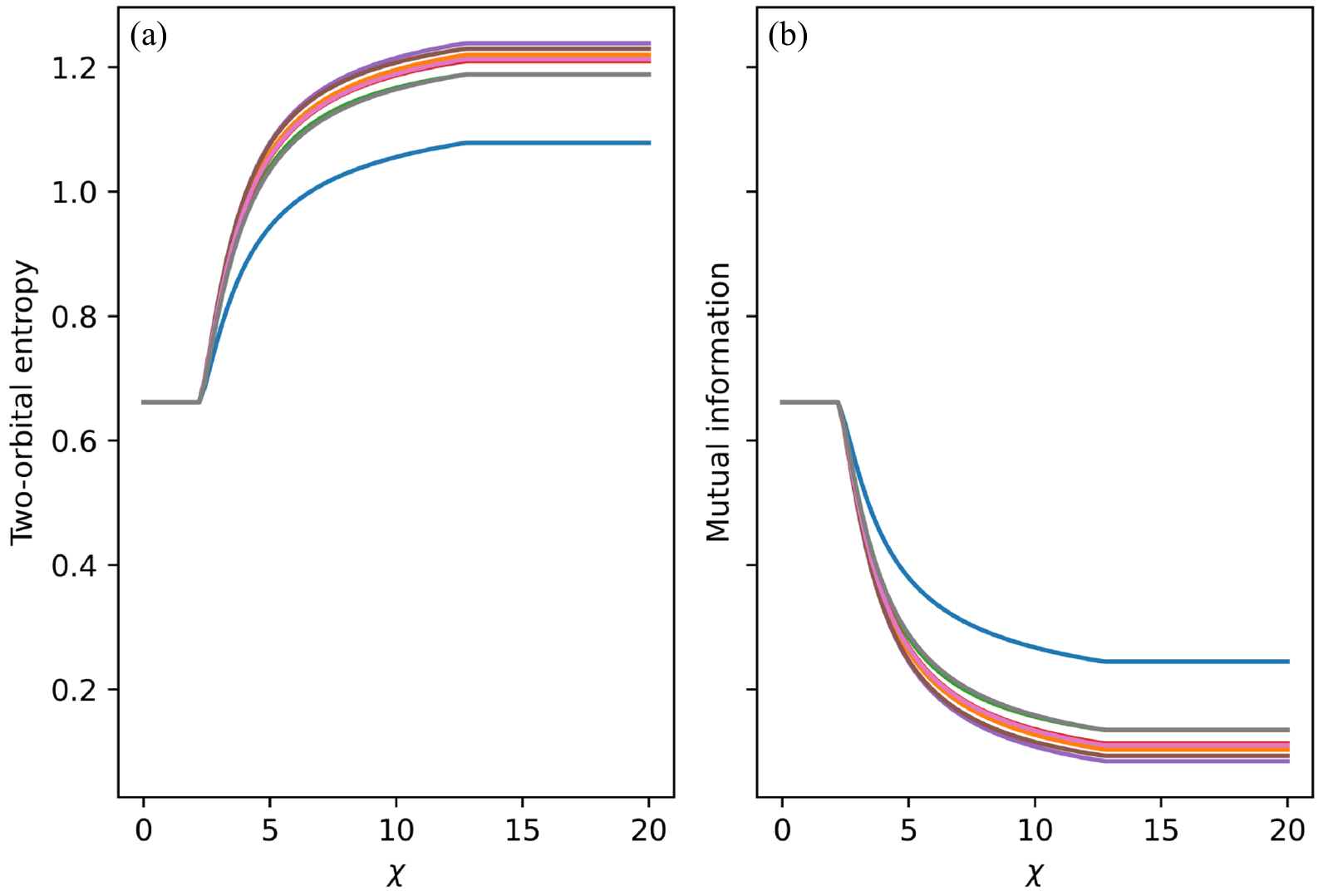}
    \caption{Same as Fig. \ref{fig:MFJ0Nj15N2} but for $N=6$.}
    \label{fig:MFJ0Nj15N6}
\end{figure}
\begin{figure}[h!]
    \centering
    \includegraphics[width=0.48\textwidth]{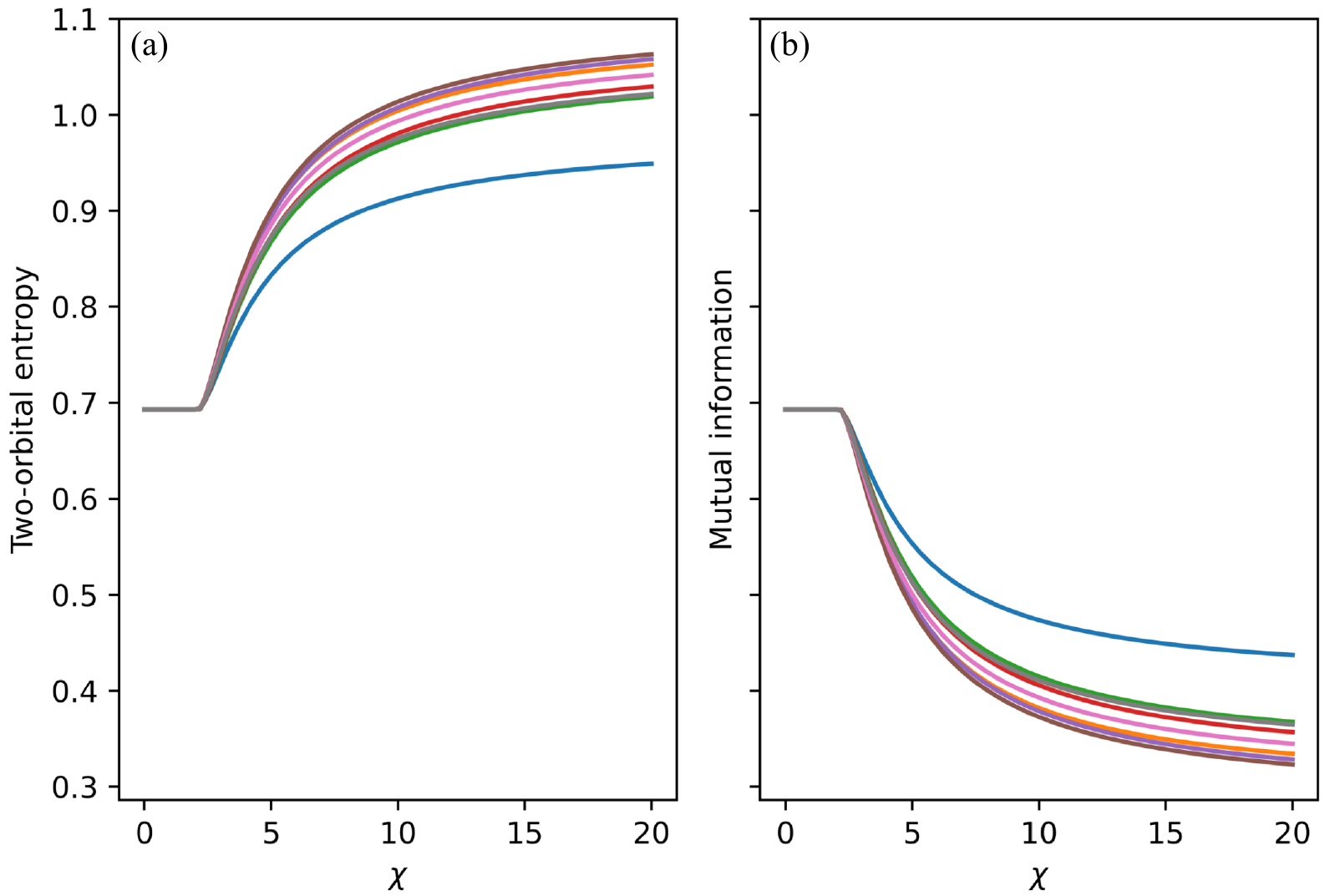}
    \caption{Same as Fig. \ref{fig:MFJ0Nj15N2} but for half-filling.}
    \label{fig:MFJ0Nj15N8}
\end{figure}

Mutual information between the $(j,m)$ and $(j,-m)$ orbitals is defined as 
\begin{equation} 
I(m,-m) = S(\rho^{(m)})+S(\rho^{(-m)})-S(\rho^{(m,-m)})
\end{equation}
Since $S(\rho^{(m)}) = S(\rho^{(-m)}) = h(\frac{\langle N \rangle}{2j+1})+h(1-\frac{\langle N \rangle}{2j+1})$ with $h(x) = -x\log x$ remains constant as a function of $\chi$ then $I(m,-m) = -S(\rho^{(m,-m)}) + C$ where $C$ is a constant. Our results indicate that, for the spherical region, the mutual information and the two orbital entropy acquire always the same non zero value. This is only possible if $I(m,-m) = S(\rho^{(m,-m)}) = S(\rho^{(m)})$.

The behavior of the the fully restored\footnote{We call `fully restored state' to the HFB state with particle number and zero angular momentum projection.} and zero angular momentum results is essentially the same, while the one of the HFB approximation may show a different behavior depending on the orbital pair and the particle number. The two orbital entropy of the fully restored state and the zero angular momentum one shows  a very similar behavior except for a global shift: the two orbital entropy of the zero angular momentum state is zero within the spherical region, while it is non zero for the fully restored state. The HFB state for this region is made of uncorrelated zero angular momentum Cooper pairs. Hence, a zero angular momentum projection doesn't change the state. However the particle number is not well defined in this case because of the non zero pairing tensor. Then, a particle number projection changes the structure of the HFB ground state, and the reduced two orbital state becomes mixed, making the two orbital entropy to be non zero. Therefore, the effect of particle number restoration is to redistribute correlations: the two orbital quantum discord is zero for all pairs since the pairing tensor is zero, but correlations between two orbital subsystems and the rest of the system (correlations among Cooper pairs) are non zero. Essentially the same happens for the deformed region regarding the angular momentum symmetry. The spontaneous symmetry breaking and restoration mechanism allows us to eliminate `fake' correlations between small subsystems (such as two orbitals) due to the mean field ansatz, and at the same time it creates correlations between bigger subsystems (among pairs of orbitals).

\section{\label{sec:conclusions}Summary, discussion and conclusions}

The spontaneous symmetry breaking of mean field solutions for fermion systems (in particular, nuclear systems) and the subsequent symmetry restoration as a mechanism to obtain an approximate ground state closer to the exact ground state of the system that is able to incorporate additional correlations beyond mean field is a fundamental topic for the theoretical nuclear physics. However, the underlying mechanisms behind the incorporation of correlations through the above mentioned mechanisms are poorly understood from a quantitative perspective. In previous work  \cite{faba_1,faba_2,faba_3} we introduced quantum information methods to quantify those correlations by considering two simple models, namely the Lipkin and Agassi models, which are characterized by abelian symmetries, concretely, the particle number and parity symmetries. In this paper, we apply these fermionic quantum information methods and tools to study a far more realistic nuclear model, the single-$j$ model, that incorporates the essential ingredients of any reasonable residual nuclear interaction and spontaneously breaks the non-abelian rotational symmetry in addition to particle number.

When dealing with rotational symmetry breaking, one has to take into account that quantum information quantities will depend on the spatial orientation of the deformed intrinsic state. In order to obtain a meaningful description of the correlation structure of the deformed state, it is necessary to remove this dependence by choosing the natural orbitals as the single particle basis in order to measure the intrinsic classical and quantum correlations within the system. Correlations can be quantified by using the following quantities: quantum discord, as a measure of the total quantum correlations between two subsystems, including entanglement and beyond; mutual information, as a measure of the total correlations between two subsystems, including quantum and classical ones; classical correlations, defined as the difference between the mutual information and the quantum discord; one body entropy, as a measure of the total correlations within the total system; and two orbital entropy, as a measure of the total correlations between two orbitals and the rest. Please note that all correlations are defined between specific partitions of the whole system.

The results for the HFB state show that the maximum amount of total correlations within the system is obtained for the spherical phase, when the total angular momentum is zero. Since the reduced state corresponding to two paired orbitals (i.e., a Cooper pair) is pure, the model's  quantum state can be described as a product of uncorrelated paired orbitals. Hence, the only contributions to the total correlations of the system are the correlations within each Cooper pair. Those can be characterized through quantum discord, mutual information and classical correlations. However, since the reduced state is pure, quantum and classical correlations are exactly the same, so the previous three quantities are essentially equivalent. A quantum phase transition (QPT) to the deformed phase can be observed as a function of $\chi$ in those quantum information measures. However, in the $N$=6 case one observes two, one corresponding to the rotational symmetry breaking and the other corresponding to the disappearance of pairing correlations, leading to particle number symmetry restoration. The two symmetries present remarkable  differences in terms of the structure of correlations of the HFB state. For the spherical phase ($\chi<\chi_c$), the total correlation of the system is maximum and the correlations within the Cooper pairs are the same for all pairs. However, for the well-defined particle number phase ($\chi>\chi_c'$ in the case of $N=6$), quantum correlations within Cooper pairs are zero, and therefore the total correlation for the whole system is zero. This means that the Cooper pairs are totally broken. When both symmetries are broken ($\chi_c<\chi<\chi_c'$) the behavior of quantum correlations within Cooper pairs depends on the specific pair. However, they generally decrease when the deformation parameter grows. This behavior is opposed to the common idea in which the mean-field state breaks the symmetry in order to incorporate more correlations, and could be a consequence of the interaction between the rotational symmetry and the particle number symmetry: when the system breaks the first one, correlations within Cooper pairs decrease, while when the system breaks the second one, correlations within Cooper pairs increase. This reveals that the interplay between symmetries of distinct nature induces a non-trivial behavior in the system's correlation structure.

When restoring the rotational symmetry by projecting to $J=0$, the mean field structure of the quantum state for the deformed phase is broken. This changes the coupling scheme of the system compared to the one corresponding to the spherical HFB state. As a consequence, we can observe that correlations within Cooper pairs are transformed to correlations among Cooper pairs. The two orbital reduced state is no longer pure and its entropy increases with $\chi$, while correlations within Cooper pairs decrease, maintaining the total correlation for the whole system constant. In the $N=6$ case, when pairing correlations are destroyed, all quantum correlations within Cooper pairs go to zero but, surprisingly, classical correlations remain (see Fig \ref{fig:MFJ0j15N6}b).

The restoration of both symmetries simultaneously through projection to $J=0$ and $N$ particles also redistributes correlations from `within pairs of orbitals' to `among pairs of orbitals'. Since the fully symmetry restored state is closer to the exact state than the HFB state in terms of energy, it is reasonable to conclude that the correlation structure for the fully symmetry restored state is closer to the structure of the exact state compared to the structure of the HFB state. Hence, the HFB state creates those `fake' correlations within Cooper pairs in order to absorb the maximum amount of complex correlations\footnote{The term complex correlations refer to correlations among bigger subsystems, for example, between a pair of orbitals and the rest of the system.} without breaking the product structure.

In this paper we have computed and analyzed the correlation structure of the HFB and symmetry restored ground states of a realistic nuclear model with abelian and non-abelian symmetries, providing some insights into the idea of the spontaneous symmetry breaking process and restoration as mechanisms that incorporate correlations within the system. This is the starting point for further work that may include the comparison of these results with those obtained for the exact ground state as obtained through the interacting shell model approach. In this respect we can mention the recent work of Ref \cite{robin2026}. Another natural extension would be to compute and compare the correlation structure for the $J\neq 0$ symmetry restored states (Yrast band). A study in this direction could give some insights on the back-bending phenomena and high-spin physics from a completely new point of view.

\acknowledgments{}

The authors want to thank the Madrid regional government, Comunidad 
Aut\'onoma de Madrid, for partial support through the project Quantum Information Technologies Madrid: 
TEC-2024/COM-84 QUITEMAD-CM.
This work has received funding from the European Union's Horizon Europe research and innovation programme under grant agreement No 101297140 (QIA-Phase2).
The  work of LMR was supported by Spanish Ministry 
of Economy and Competitiveness (MINECO) Grants No. PID2024-159559NB-C21.

\appendix


\section{\label{sec:AppendixA}The two-orbital reduced density matrix of paired orbitals in the natural basis is pure for a HFB state}

The two-orbital reduced density matrix between orbitals $i$ and $j$ for a general valid fermionic state (this is, a state that satisfies PSSR) \cite{Friis_2016} is given by
\begin{widetext}
\begin{equation}
\rho^{(i,j)} = 
\begin{pmatrix}
1-\gamma_{i,i}-\gamma_{j,j}+\gamma_{i,j,i,j} & 0 & 0 & \kappa_{j,i}^* \\
0 & \gamma_{j,j}-\gamma_{i,j,i,j} & \gamma_{j,i} & 0 \\
0 & \gamma_{j,i}^* & \gamma_{i,i}-\gamma_{i,j,i,j} & 0 \\
\kappa_{j,i} & 0 & 0 & \gamma_{i,j,i,j} \\
\end{pmatrix}
\end{equation}
\end{widetext}
where $\gamma_{i,j} = \langle c^\dagger_jc_i\rangle$ are the elements of the one-body density matrix,  $\gamma_{i,j,i,j} = \langle c^\dagger_ic^\dagger_jc_jc_i\rangle$ are the elements of the two-body density matrix and $\kappa_{i,j} = \langle c_jc_i\rangle$ are the elements of the pairing tensor.
Taking into account that the quantum state is a HFB state, the two-body density matrix factorizes and can be written as 
\begin{equation}
\gamma_{i,j,i,j} = \gamma_{i,i}\gamma_{j,j} + \Delta_{i,j}
\end{equation}
with $\Delta_{i,j} = \abs{\kappa_{i,j}}^2-\abs{\gamma_{i,j}}^2$. Moreover, if the orbital basis is the natural one, the one-body density matrix will be always diagonal by definition \cite{RingSchuck1980}. On the other hand, if we set $i$ and $j$ to a couple of paired natural orbitals, then $\gamma_{i,i} = \gamma_{j,j}$. This can be demonstrated using the Bloch-Messiah theorem. Taking into account those conditions, the two-orbital reduced density matrix can be written as
\begin{widetext}
\begin{equation}
\label{eq:rho_AB}
\rho^{(i,j)} = 
\begin{pmatrix}
(1-\gamma_{i,i})^2+\abs{\kappa_{i,j}}^2 & 0 & 0 & \kappa_{j,i}^* \\
0 & (1-\gamma_{i,i})\gamma_{i,i}-\abs{\kappa_{i,j}}^2 & 0 & 0 \\
0 & 0 & \gamma_{i,i}(1-\gamma_{i,i})-\abs{\kappa_{i,j}}^2 & 0 \\
\kappa_{j,i} & 0 & 0 & \gamma_{i,i}^2+\abs{\kappa_{i,j}}^2 \\
\end{pmatrix}
\end{equation}
\end{widetext}
Since the quantum state is a HFB state, the one-body density and the pairing tensor are not completely independent. They fulfill the following condition \cite{RingSchuck1980}
\begin{equation}
    \gamma^2-\gamma = -\kappa\kappa^\dagger
\end{equation}
In the natural orbital basis, it is reduced to 
\begin{equation}
\label{eq:HFB_rho_kappa_condition}
    \gamma_{i,i}(1-\gamma_{i,i})-\abs{\kappa_{i,j}}^2 = 0
\end{equation}
which is exactly the value of the diagonal matrix elements of the innermost $2\times 2$ matrix of Eq. (\ref{eq:rho_AB}). Now we diagonalize the outermost $2\times 2$ matrix. We obtain the following equation for the eigenvalues
\begin{widetext}
\begin{equation}
\label{eq:outer_eigvals_rhoAB}
    \lambda^2 - \lambda\bigg(\gamma_{i,i}^2+(1-\gamma_{i,i})^2+2\abs{\kappa_{i,j}}^2\bigg) + \bigg( (1-\gamma_{i,i})^2 +\abs{\kappa_{i,j}}^2\bigg)\bigg( \gamma_{i,i}^2 +\abs{\kappa_{i,j}}^2\bigg)-\abs{\kappa_{i,j}}^2 = 0
\end{equation}
\end{widetext}
Since the innermost diagonal block is equal to zero, the trace of $\rho^{(i,j)}$ is given by 
\begin{equation}
\label{eq:inner_trace_rhoAB}
    \Tr(\rho^{(i,j)}) = \gamma_{i,i}^2+(1-\gamma_{i,i})^2+2\abs{\kappa_{i,j}}^2
\end{equation}
Using the condition $\Tr(\rho^{(i,j)}) = 1$ together with Eq. (\ref{eq:inner_trace_rhoAB}), and the relation between the one-body density and the pairing tensor of Eq. (\ref{eq:HFB_rho_kappa_condition}), Eq. (\ref{eq:outer_eigvals_rhoAB}) reduces to $\lambda^2-\lambda = 0$. Hence, the eigenvalues of the two-orbital reduced density matrix for a couple of paired natural orbitals of a HFB state are zero or one, so the reduced state is always pure.

\section{\label{sec:AppendixB}Elements of the one body density matrix for a $J=0$ state in the single-$j$ model}

The fermionic operator that creates a particle with angular momentum $(j,m)$, $c^\dagger_m$, is the $m$-th spherical component of a irreducible tensor operator of rank $j$. As a consequence
\begin{equation}
    R(\Omega)c^\dagger_mR^\dagger(\Omega) = \sum_{m'}D^{(j)}_{m'm}(\Omega)c^\dagger_{m'}
\end{equation}
which is the transformation law under rotations for the $m$-th spherical component of a rank-$j$ irreducible tensor operator. We can use this property, together with Eq. \ref{eq:P_operator_J0} and the composition equation for Wigner matrices,
\begin{equation}
\begin{split}
    &D^{(j_1)}_{mm'}(\Omega)D^{(j_2)}_{nn'}(\Omega) = \sum_{J' = \abs{j_1-j_2}}^{j_1+j_2}\bra{j_1m;j_2n}\ket{J',m+n}\\ &\bra{j_1m';j_2n'}\ket{J',m'+n'}D^{(J')}_{m+n,m'+n'}(\Omega)
\end{split}
\end{equation}
to compute the matrix elements of the one body density matrix for a $J=0$ projected mean-field state:
\begin{widetext}
\begin{equation}
\label{eq:onebody_zeroangularmom_long}
\begin{split}
    &\bra{\phi}P^\dagger c^\dagger_mc_nP\ket{\phi} = \bra{\phi}\frac{1}{8\pi^2}\int d\Omega R^\dagger(\Omega)c^\dagger_mc_nP\ket{\phi}
    = \bra{\phi}\frac{1}{8\pi^2}\int d\Omega\sum_{m'n'}D^{(j)*}_{mm'}(\Omega)D^{(j)}_{nn'}(\Omega)c^\dagger_{m'}c_{n'}R^\dagger(\Omega)P\ket{\phi} \\ 
    &= \bra{\phi}\frac{1}{8\pi^2}\int d\Omega\sum_{m'n'}(-1)^{m-m'}\sum_{J'=0}^{2j}\bra{j-m;jn}\ket{J',n-m}\bra{j-m';jn'}\ket{J',n'-m'} 
    D^{(J')}_{n-m,n'-m'}(\Omega)c^\dagger_{m'}c_{n'}R^\dagger(\Omega)P\ket{\phi} \\
    &= \sum_{m'n'}(-1)^{m-m'}\sum_{J'=0}^{2j}\frac{1}{2J'+1}\bra{j-m;jn}\ket{J',n-m}\bra{j-m';jn'}\ket{J',n'-m'}
    \bra{\phi}c^\dagger_{m'}c_{n'}P^{J'\dagger}_{n-m,n'-m'}P\ket{\phi} \\
    &= \delta_{n,m}\sum_{m'}(-1)^{m-m'}\bra{j-m;jm}\ket{0,0}\bra{j-m';jm'}\ket{0,0}\bra{\phi}c^\dagger_{m'}c_{m'}P\ket{\phi} 
    = \delta_{n,m} \frac{1}{2j+1}\bra{\phi}\hat{N}P\ket{\phi} \\
\end{split}
\end{equation}
\end{widetext}
where we have used the time reversal property $D^{(j)*}_{mm'}(\Omega) = (-1)^{m-m'}D^{(j)}_{-m-m'}(\Omega)$, $P^{J'\dagger}_{n-m,n'-m'}P = P\delta_{J',0}\delta_{n,m}\delta_{n',m'}$ and $\bra{j,-m;j,m}\ket{0,0} = \frac{(-1)^{j+m}}{\sqrt{2j+1}}$.

\bibliography{bibliography} 
\end{document}